# The Structure-tuning Magnetism in the $Co(Nb_xTa_{1-x})_2O_6$ Series

Sirui Huang[1, #], Lun Jin[1, #, *], Yuchen Liu[1], Xiyu Chen[1] and Leili Tan[1]

[1]Key Laboratory of Quantum Materials and Devices of Ministry of Education, School of Physics, Southeast University, Nanjing 211189, China

# These authors contributed equally to this work.

* E-mails of corresponding authors: ljin@seu.edu.cn

**Abstract:** Low dimensional quantum materials have been extensively studied within the physics community, among which $Co^{2+}$ is one of the most popular magnetic ions to be embedded into the structural motif. $CoNb_2O_6$ and $CoTa_2O_6$ both have adopted an effective spin-1/2 state, with the former attracting enduring interest while the latter being rarely investigated. This difference can be attributed to their distinct crystal structures, not only in crystal systems, but more important, how the $CoO_6$ octahedra are connected. We successfully merge these two compounds together and map out the doping limits. We find out Ta and Nb dopants modify the properties of the parent compounds in subtle but different ways and construct a magnetic phase diagram of the $Co(Nb_xTa_{1-x})_2O_6$ ($0 \le x \le 1$) series. In addition, our results suggest that Ta-doping may pave a way to strengthen the lower-dimensional quantum phenomena in $CoNb_2O_6$ before its crystal structure starts to melt down.

## I. INTRODUCTION

Low dimensional quantum materials have attracted enduring interest in condensed-matter and materials physics, owing to their exotic phenomena ranging from complex ordered magnetic ground states to frustrated magnetism, quantum fluctuations etc.[1–7]. Among abundant choices in the periodic table, $Co^{2+}$ ($3d^7$) stands out fairly easily due to its Kramers' ion nature and strong single-ion anisotropy[8–12]. Thus, by putting it into a low-dimensional structural motif, the spins in $Co^{2+}$ tend to align themselves along the easy axis and adopt an effective spin-1/2 state that allows them to communicate via Ising interactions[13–19]. Cobalt-based quasi-1-dimensional materials are of special interest because their dimensionality (i.e., the relative strengths of intra-chain and inter-chain magnetic couplings) can be tuned over a wide range. There are fruitful choices for not only how to construct these Co-containing 1D chains (edge- or face-sharing, zig-zag or screw type etc.), but also the way of bridging them together (through tetrahedra, octahedra or even hybrid units)[20].

$CoNb_2O_6$, as the most famous member of the columbite family, has been under extensive investigation for decades. This is because it can be considered as an ideal system to study the interplay of quantum fluctuations and quantum criticality [14, 21, 22], which are currently two of the most intriguing topics in condensed-matter physics. However, studies regarding the modifications of this classic material rarely come up. Though its derivative $CoTa_2O_6$ has also been studied, but not to a comparable extent so far[23–26], it does provide a potential perspective to further manipulate the dimensionality of $CoNb_2O_6$, hence its quantum phenomena.

$CoNb_2O_6$ crystallizes in an orthorhombic unit cell (space group *Pbcn* (# 60))[27], while its severely less studied sister compound $CoTa_2O_6$ crystallizes in a tetragonal unit cell (space group $P4_2/mnm$ (# 136))[23]. Despite the fact that their unit cells belong to different crystal systems, what makes a real difference in getting attention from the physics community is how their $CoO_6$ octahedra are connected. In $CoNb_2O_6$, it consists of zig-zag chains running along the *c* axis that are made of edge-sharing oxygen octahedra, with two adjacent $CoO_6$ 1D chains separated by two $NbO_6$ 1D chains (Figure 1a). This arrangement yields in distinct intra- and inter-chain magnetic couplings. The competition between them further leads to complicated ordered and/or disordered states and other emergent physical phenomena that remain under active discussions up to today[22, 28]. In contrast, $CoTa_2O_6$ adopts a cation-ordered rutile-derived structure (Figure 1b). It results in a Co-Ta-Ta-Co sequence within each 1D chain, which might dilute the intra-chain Co-Co magnetic coupling to a certain extent and forfeit some exotic phenomena[23, 24].

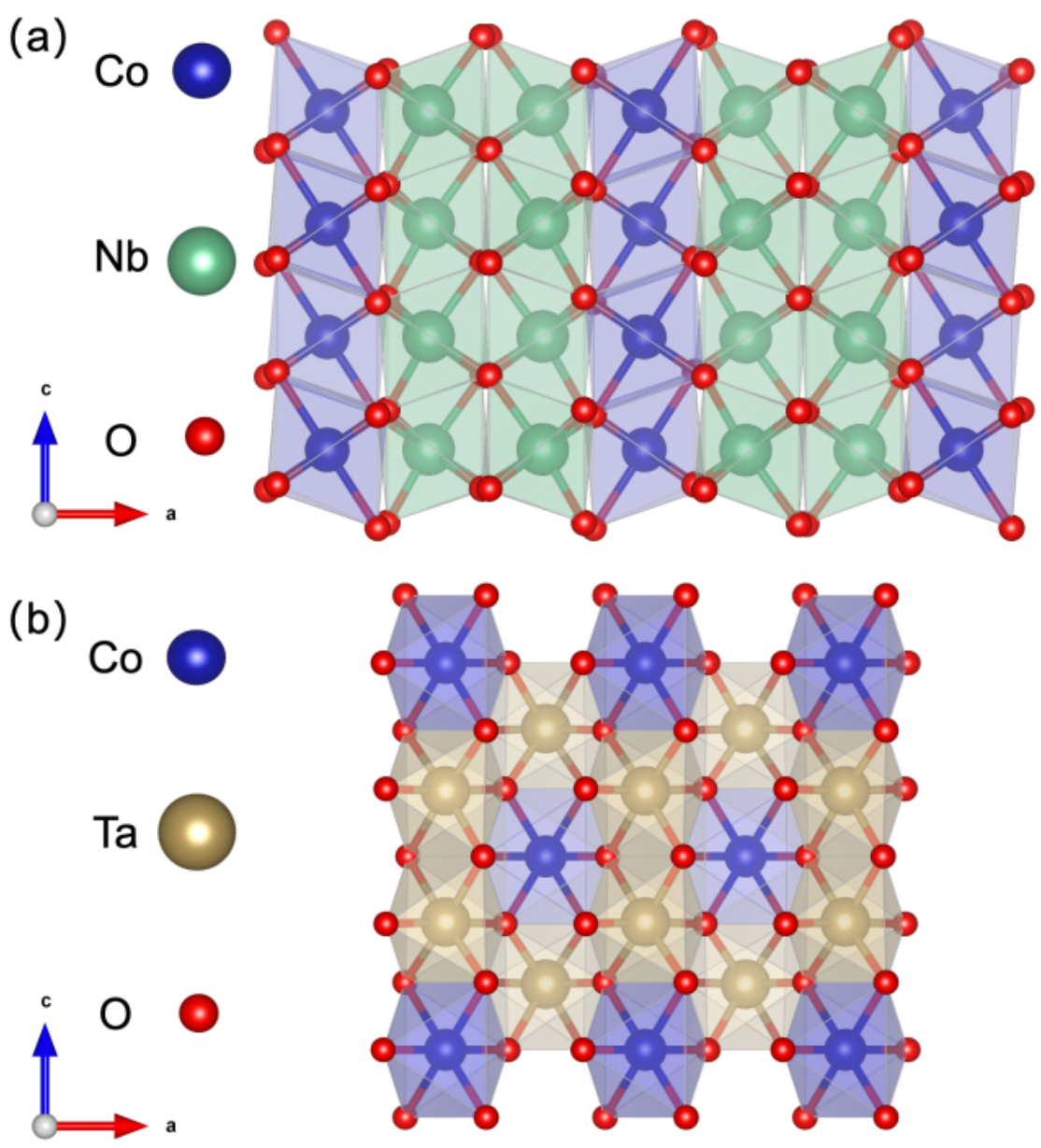

FIG. 1. The crystal structure of (a) $CoNb_2O_6$ and (b) $CoTa_2O_6$, with visualization of 1D chains consisted of edge-sharing $CoO_6$ and $(Nb/Ta)O_6$.

Due to the fact that $CoNb_2O_6$ is structurally distinct from its sister compound $CoTa_2O_6$, solvating one into the other[29] to explore the impact on physical properties was rarely studied. Therefore, in this work, we took attempts to merge these two compounds with different ratios and mapped out the doping limits. Hence, we examined how these dopants modify the properties of parent compounds and construct a magnetic phase diagram of the $Co(Nb_xTa_{1-x})_2O_6$ ($0 \le x \le 1$) series. Last but not least, our results suggest that Ta-doping may pave a way to strengthen the lower-dimensional quantum phenomena in $CoNb_2O_6$, before the ideal quasi-one-dimensional crystal structure starts to melt down.

## II. METHODS

### A. Materials Synthesis

The polycrystalline samples of the $Co(Nb_xTa_{1-x})_2O_6$ ($0 \le x \le 1$) series were synthesized by solid-state reactions. Stoichiometric amounts of CoO (Aladdin, 99.94%), $Nb_2O_5$ (Macklin, 99.99%), and $Ta_2O_5$ (Macklin, 99.99%) powders were thoroughly mixed and ground using an agate mortar and pestle, and then transferred into an alumina crucible. The reaction mixtures were heated in air to 1300 °C at a ramp rate of 4 °C/min and held for 12 hours. Three to four heating cycles under the same conditions with intermittent grindings were taken to ensure the complete reactions and phase homogeneity. After the final sintering, the samples were furnace-cooled to room temperature. For heat capacity measurements, the samples were pelletized using a uniaxial press with a 1/8-inch diameter die under a load of 430 kg for 10 minutes. The pellets were subsequently sintered at 1300 °C for 12 hours in air to improve mechanical stability and ensure good thermal contact during measurements.

### B. Structural Characterization

The phase purity and crystal structure of the $Co(Nb_xTa_{1-x})_2O_6$ ($0 \le x \le 1$) samples were characterized by powder X-ray diffraction at room temperature using a Rigaku diffractometer with Cu Kα radiation. Diffraction patterns were collected over the $2\theta$ range between 10° and 70° with a step size of 0.02°. The obtained laboratory p-XRD data were analyzed by Rietveld refinement using the GSAS-II software package.

### C. Physical Property Characterizations

The magnetization data were collected using the VSM option of a Quantum Design Physical Property Measurement System. Temperature-dependent magnetization ($M$) data were collected under applied external fields ($H$) of 100, 300, 500, 700, and 1000 Oe over the temperature range of 1.8 to 300 K. Magnetic susceptibility is defined as $M/H$. The measurements were performed on powder samples. Heat capacity was measured using the standard relaxation method in the Physical Property Measurement

System over the temperature range of 1.8 to 100 K under zero magnetic field with an appropriate range used for background phonon fitting. The measurements were carried out on pelletized samples with sintering conditions mentioned above.

## III. RESULTS

### A. Structural Characterization

Although Ta sits in one slot below Nb in the periodic table, they do offer comparable ionic radii while adopt the oxidation state of +5 combined with an octahedral coordination environment[30]. However, both end members of the $Co(Nb_xTa_{1-x})_2O_6$ series have different crystal structures, as illustrated in Figure 1. Thus, the limit of solvating $CoNb_2O_6$ into $CoTa_2O_6$, and to pinpoint the potential structural shift needs to be carefully explored.

In this work, the initial synthesis attempt of the $Co(Nb_xTa_{1-x})_2O_6$ series was carried out in 25% intervals. The phase purity was examined by the powder XRD patterns collected on polycrystalline samples and stacked vertically for the comparison (Figure 2a). In the Ta-rich side, the crystal structure remains as tetragonal, while in the Nb-rich side, it adopts an orthorhombic structure, all consistent with both end members. It is found that the pattern of the $x = 0.5$ sample contains peaks corresponding to the Bragg positions of both end members, indicating it as a two-phase mixture. From the initial synthesis attempt, we conclude that $CoNb_2O_6$ and $CoTa_2O_6$ can solvate into each other to a certain extent, accompanied by the conservation of the crystal structure adopted by each end member. Therefore, we then undertook a much smaller interval (10%) in composition and constructed a structural phase diagram of the $Co(Nb_xTa_{1-x})_2O_6$ series (Figure 2b). Rietveld refinements were performed against the collected XRD pattern of each composition, with extra attention paid to whether it is a single-phase compound or not (Figure S1). In the Ta-rich side ($0 \leq x \leq 0.4$, the green area in Figure 2b) and the Nb-rich side ($0.75 \leq x \leq 1$, the pink area in Figure 2b), the lattice parameters remain almost unchanged, respectively (Table S1), in a good agreement with the comparable ionic radii of $Ta^{5+}$ and $Nb^{5+}$ in an octahedron. The compounds in the range of $0.4 < x <$ 0.75 (the grey area in Figure 2b) are clearly two-phase, with the fitting statistics much improved by adding a secondary phase during the refinements.

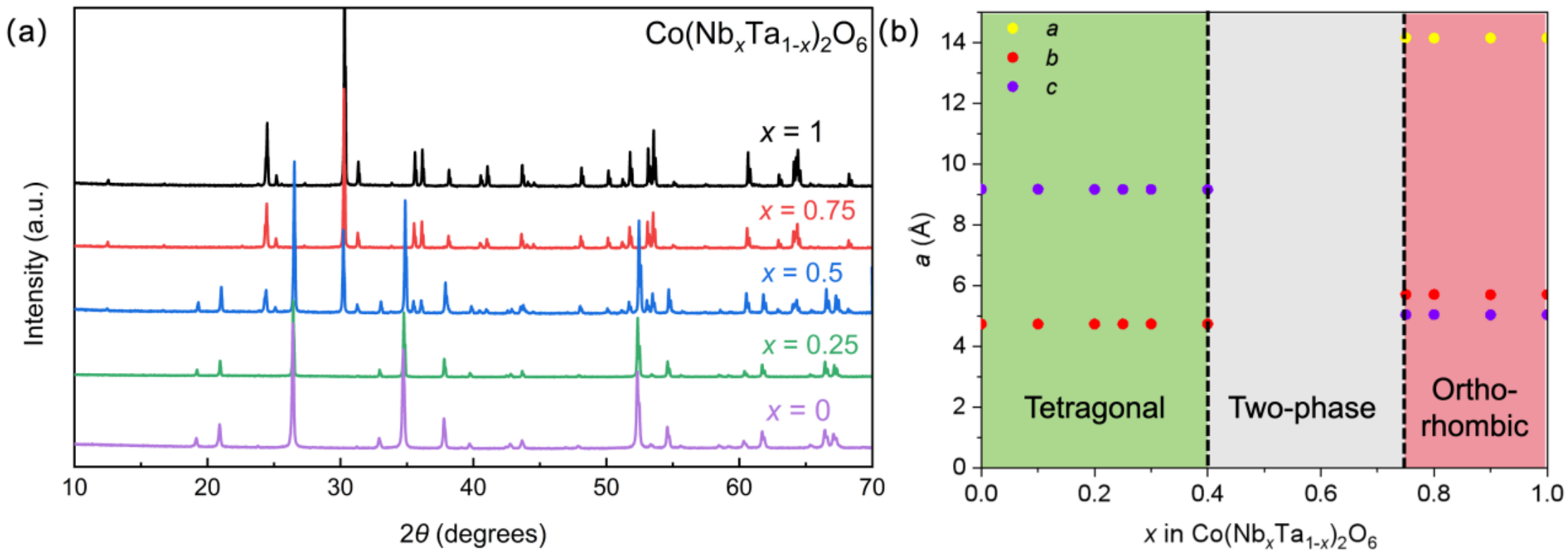

FIG. 2. (a) The stacked XRD patterns of selected compositions of the Co(Nb$_x$Ta$_{1-x}$)$_2$O$_6$ series. (b) The structural phase diagram constructed for the Co(Nb$_x$Ta$_{1-x}$)$_2$O$_6$ series, with the lattice parameters refined from the collected XRD patterns.

**B. Magnetism**

After studying the magnetism of both end members, we notice that $CoNb_2O_6$ and $CoTa_2O_6$ behave differently under applied magnetic fields, as a result of their distinct crystal structures, i.e., how those Co-containing 1D chains are constituted and joint together. The zero-field-cooled (ZFC) temperature-dependent magnetization data were collected from the Co(Nb$_x$Ta$_{1-x}$)$_2$O$_6$ series, under selected magnetic fields. In our case, $CoTa_2O_6$ only has one at $T_N$ ~ 6.5 K, while $CoNb_2O_6$ has two intrinsic magnetic transitions, at $T_{N1}$ ~ 2.7 K and $T_{N2}$ ~ 2.0 K under $H$ = 100 Oe, both in a good agreement with previous studies[26, 31]. In addition, the magnetic susceptibility of $CoTa_2O_6$ (Figure S2) is two orders of magnitude smaller than that of the Nb analogue (Figure S3), consistent with the dilution of the Co-Co magnetic coupling by Ta in its cation-ordered rutile-derived structure (Figure 1b). For compositions in between, detailed results are provided in the following.

In the Ta-rich side ($0 \leq x \leq 0.4$), the magnetic susceptibility $\chi$ plotted against temperature $T$ of these compositions in general obey the trend of the end member $CoTa_2O_6$ (Figure 3a). However, by taking a closer look at the d$\chi$/d$T$ data in the low-temperature regime, we find that Nb-for-Ta substitution gradually and subtly shifts the transition temperature $T_N$ from ~ 6.5 K to lower temperatures (Figure 3b). The magnetic susceptibility $\chi$ of these samples were measured under various fields and they appear to be insensitive to external fields (Figure S4-S9), in contrast with the Nb-rich side samples discussed below. The magnetic susceptibility data, over a suitable temperature range (selected as the straight-line part of the $1/\chi$ vs. $T$ curves, marked in red in Figure 3c) for each composition were fitted to the Curie-Weiss law ($\chi = C/(T - \theta) + \chi_0$), to yield the Curie constants $C$, Weiss temperatures $\theta$ and effective moments $\mu_{eff}$. The results of $CoTa_2O_6$ are showing here as a representative (Figure 3c). The Curie constant $C$ (3.0421(2) $cm^3$ K $mol^{-1}$) leads to an effective moment of $\mu_{eff}$ = 4.93 $\mu_B$/f.u. (calculated based on the equation $\mu_{\mathrm{eff}} = \sqrt{8C}\mu_{\mathrm{B}}$). This value is larger than the spin-only expectation for $S$ = 3/2 $Co^{2+}$ (high-spin $d^7$, 3.87 $\mu_B$/f.u.), revealing that spin-orbit coupling substantially contributes to the magnetism in $CoTa_2O_6$. The Weiss temperature $\theta$ is determined to be -32.27(2) K, indicating the dominant antiferromagnetic coupling between $Co^{2+}$ centers. The C-W fitting results of the rest compositions are presented in Figure S2-S3 and Table S2, showing acceptable fluctuations in the values of Curie constants $C$, Weiss temperatures $\theta$ and effective moments $\mu_{eff}$.

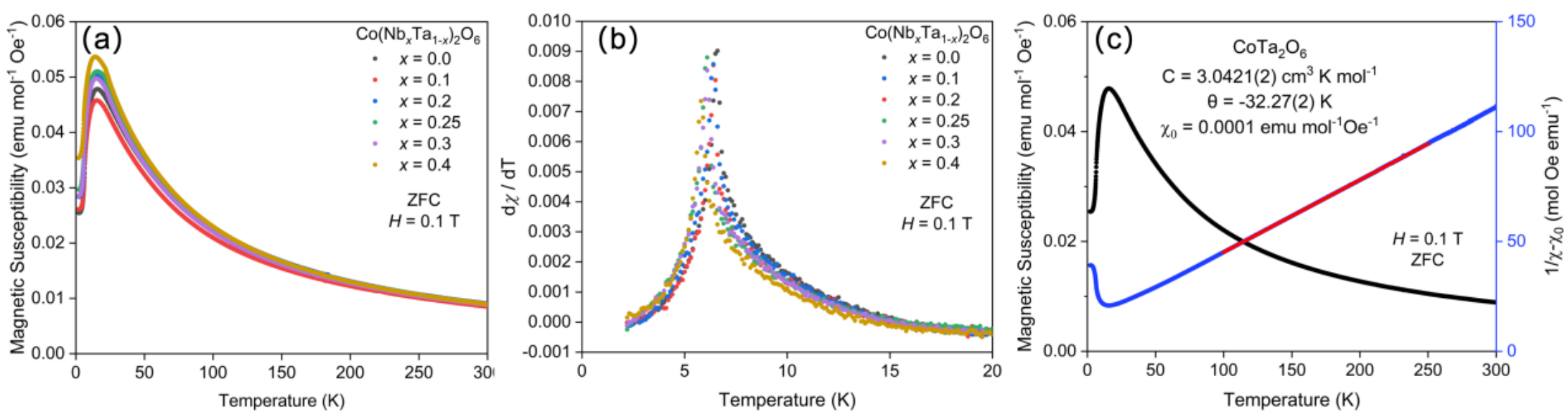

FIG. 3. (a) The temperature-dependent magnetization data collected from the Ta-rich side compositions ($0 \le x \le 0.4$) of the $Co(Nb_xTa_{1-x})_2O_6$ series. (b) The $d\chi/dT$ data plotted against temperature $T$ in the zoom-in low temperature area to visualize $T_N$. (c) The Curie-Weiss fit results of $CoTa_2O_6$.

In $CoNb_2O_6$, it is commonly acknowledged that $T_{N1}$ ~ 2.7 K is associated with a long-range 3D magnetic ordered phase while $T_{N2}$ ~ 2.0 K has a pronounced 2D character.[27, 32–34] Thus, we measured the magnetic susceptibility $\chi$ for the Nb-rich side ($0.75 \leq x \leq 1$) samples under $H$ = 100, 300, 500, 700 and 1000 Oe in the low-temperature regime. We are interested in seeing how the Ta-doping potentially affect intrinsic magnetic transitions $T_{N1}$ and $T_{N2}$, especially their sensitivities to the external fields. In general, by plotting the magnetic susceptibility $\chi$ (Figure 4a) as well as $d\chi/dT$ (Figure S10-S13) against temperature $T$ of these compositions, we find that $T_{N1}$ and $T_{N2}$ do not shift significantly, in contrary to the Ta-rich side samples. However, they do reveal something more subtle instead.

For all the measured compositions, the shorter-range $T_{N2} \sim 2.0$ K with a pronounced 2D character can only be observed under $H$ = 100 and 300 Oe, while it is suppressed under $H \geq 500$ Oe in each sample (Figure S10-S13). The long-range 3D magnetic ordering at $T_{N1} \sim 2.7$ K is shown to be severely interfered by Ta-doping, as reflected by the gradual deform of the peak shape in the $d\chi/dT$ plots. The peak practically becomes an anomaly in the $x$ = 0.8 sample, and disappear in the $x$ = 0.75 sample, which is on the edge of the Nb-rich side in our study. Data collected under $H$ = 700 Oe for each composition are showing in Figure 4b-4e to illustrate this subtle but continuous change as a representative, while the rest can be found in Figure S10-S13. In addition, the C-W fitting results of compositions in the Nb-rich side are also presented in Table S2, analyzed in a consistent way as those in the Ta-rich side. A noticeable increase in the values of Curie constants $C$, hence effective moments $\mu_{eff}$, is clearly evidenced. This suggests that the spin-orbit coupling of $Co^{2+}$ contributed to the magnetism in a more substantial way for compositions in the Nb-rich side compared to the Ta-rich side, which could be largely attributed to their distinct crystal structures.

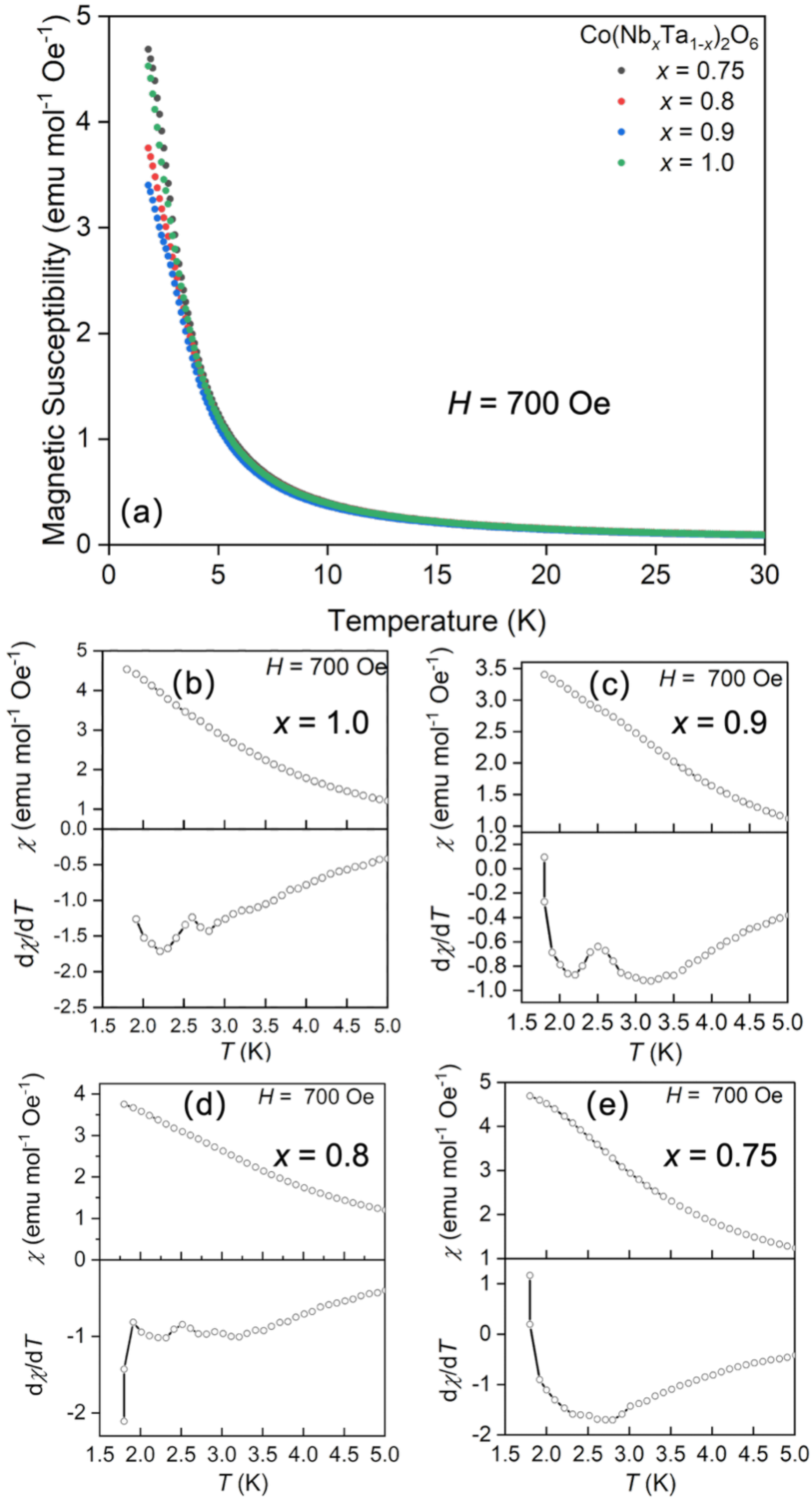

FIG. 4. (a) The temperature-dependent magnetization data collected from the Nb-rich side compositions ($0.75 \leq x \leq 1$) of the $Co(Nb_xTa_{1-x})_2O_6$ series. (b-e) The $d\chi/dT$ data plotted against temperature $T$ in the zoom-in low temperature area to visualize the continuous change in $T_{N1}$ under 700 Oe.

### C. Heat Capacity

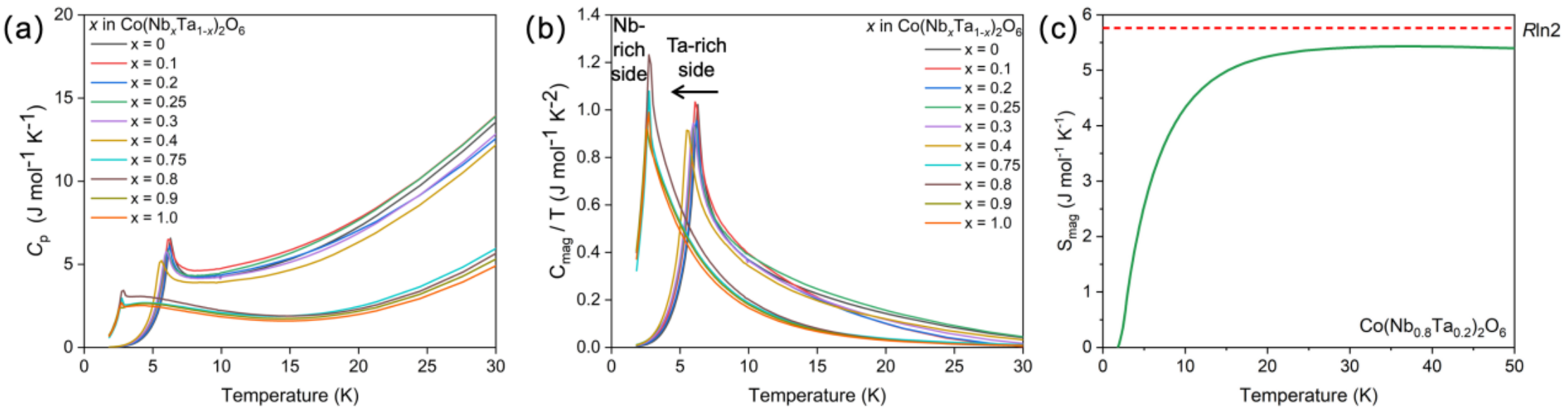

FIG. 5. (a) The total heat capacity data $C_p$ and (b) the magnetic contribution $C_{mag}/T$ plotted against temperature $T$ for all the compositions of the $Co(Nb_xTa_{1-x})_2O_6$ series made in this study. (c) The magnetic entropy $S_{mag}$ plotted against temperature $T$ for $Co(Nb_{0.8}Ta_{0.2})_2O_6$ as a representative.

In order to further verify the observed transitions in magnetic data, heat capacity data were collected from the sintered pellet of each composition of the $Co(Nb_xTa_{1-x})_2O_6$ series made in this study under zero field from 1.8 K to 100 K. The total heat capacity data $C_p$ are plotted against temperature up to 30 K in Figure 5a for better visualization of the features in the low-temperature regime. All samples are too resistive to be measured at ambient temperature and pressure, meaning there are no conduction electrons present that would yield a significant $C_{electron}$ contribution. Thus, the magnetic contribution ($C_{mag}$) to the total heat capacity ($C_p$) can be isolated by subtracting the fitted lattice contribution ($C_{lat}$) from the total heat capacity data ($C_p$).

The heat capacity data over the temperature range that is well above the region for magnetic transitions can be fitted to extract the lattice contribution ($C_{lat}$) using the following equation[35],

$$C_{lat}(T) = \sum_{i=1}^{3} n_i C_D(\Theta_{Di}, T)$$

Where $C_D(\Theta_D, T)$ is the Debye heat-capacity function,

$$C_D(\Theta_D, T) = 9R(\frac{T}{\Theta_D})^3 \int_0^{\frac{\Theta_D}{T}} \frac{x^4 e^x}{(e^x - 1)^2} dx$$

$R$ is the ideal gas constant, and $\Theta_{D1}$, $\Theta_{D2}$ and $\Theta_{D3}$ represent the effective Debye temperatures. To ensure the validity of the fitting, the total number of effective oscillators is strictly constrained to be equal to the total number of atoms in the formula unit $\sum_{i=1}^{3} n_i = 9$. The detailed fitting parameters are listed in Table S3. The yielded magnetic contribution $C_{mag}$ is then divided by temperature T so that the magnetic entropy change $\Delta S_{mag}$ can be estimated by taking the integral of $C_{mag}/T$ over the measured temperature range.

The yielded magnetic contribution $C_{mag}/T$ are plotted against temperature $T$ in Figure 5b. The peak observed in the curves for compositions in Ta-rich side ($0 \leq x \leq 0.4$) shows a clear and continuous shift towards lower temperatures, consistent with the trend shown in the $\chi$-$T$ curves and aligning perfectly with the long-range nature of this AFM ordering. In contrast, the peak observed in the curves for compositions in Nb-rich side ($0.75 \leq x \leq 1$) remain more or less unshifted around $T_{N1} \sim 2.7$ K, in a good agreement with the above-mentioned results in magnetism and the reported long-range nature of this transition in literature. An effective $S = ½$ ground state of $Co^{2+}$ (high-spin $3d^7$ configuration) is verified for compositions in Ta-rich and Nb-rich sides, consistent with both end members[13, 24, 31, 34]. The results for $Co(Nb_{0.8}Ta_{0.2})_2O_6$ are shown in Figure 5c as a representative, while those of the rest compositions can be found in Figure S18-S19. In conclusion, we find that the experimental saturation value of $\Delta S_{mag}$ approaches the Ising spin prediction ($R$ ln (2) = 5.76 J/mol/K) rather than the Heisenberg spin prediction for an $S = 3/2$ system ($R$ ln $(2S + 1)$ = 11.53 J/mol/K).

In addition, we notice that in Nb-rich side, the broad hump just above $T_{N1} \sim 2.7$ K gets enhanced in $x = 0.8$ sample (Figure 5a). This hump is reported to be associated with the shorter-range, lower-dimensional magnetic contributions rather than the long-range 3D ordering[34, 36]. Therefore, we postulate that the enhanced broad hump observed in $x = 0.8$ sample might be attributed to the strengthened lower-dimensional quantum phenomena due to Ta-doping, under the circumstance that the crystal structure of $CoNb_2O_6$ is perfectly conserved. Further Ta-doping could result in the perturbation of shorter-range Co-Co interactions, or even the collapse of the ideal crystal structure, i.e., transit to the cation-ordered rutile-derived structure of $CoTa_2O_6$. Besides, the absence of this broad hump in all the compositions studied in Ta-rich side ($0 \leq x \leq 0.4$) could further validate our argument.

## IV. DISCUSSION

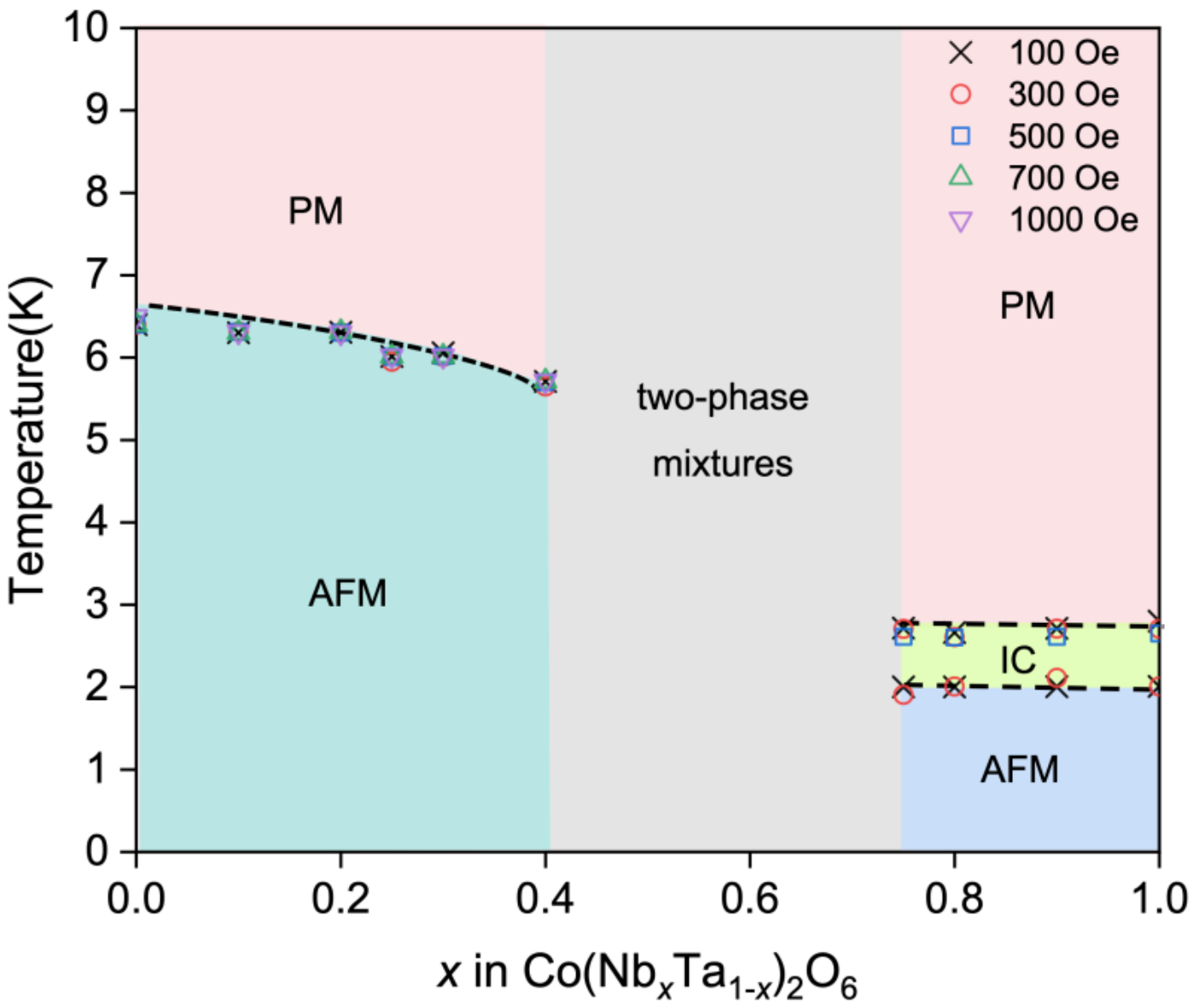

FIG. 6. The magnetic phase diagram of the $Co(Nb_xTa_{1-x})_2O_6$ series. PM stands for paramagnetic state, AFM stands for antiferromagnetic state, IC stands for incommensurate state.

Based on our results presented above, we construct the magnetic phase diagram of the $Co(Nb_xTa_{1-x})_2O_6$ series (Figure 6). Although $Ta^{5+}$ and $Nb^{5+}$ have comparable ionic radii in an octahedron[30], the distinctly different crystal structures of $CoTa_2O_6$ and $CoNb_2O_6$ put limits on the solvation of one into the other, as denoted by the grey-shaded area (two-phase mixtures) in Figure 6. In Ta-rich side ($0 \leq x \leq 0.4$), there is only one PM-to-AFM transition observed for each composition and $T_N$ is gradually lowered with respect to increasing Nb dopants. This transition is confirmed to be long-range by the λ-shape peak in heat capacity curves. The Nb dopants cannot introduce any magnetic features of $CoNb_2O_6$ into the system before reaching the upper doping limit $x = 0.4$. In contrast, the two intrinsic transitions ($T_{N1} \sim 2.7$ K and $T_{N2} \sim 2.0$ K) of compositions in Nb-rich side do not shift in general, instead, they become easier to be suppressed by the external field with elevated Ta doping concentration, especially the long-range one ($T_{N1}$

~ 2.7 K). Combined with the heat capacity results, it suggests that Ta-doping may severely interfere the 3D magnetic ordering and hence strengthen the lower-dimensional quantum phenomena in $CoNb_2O_6$, before it melts down the ideal quasi-one-dimensional crystal structure.

## V. CONCLUSIONS

In this work, we have successfully made a series of polycrystalline samples of $Co(Nb_xTa_{1-x})_2O_6$. We find out that despite the distinctly different crystal structures of $CoTa_2O_6$ and $CoNb_2O_6$, one does solvate into the other to a certain extent. In Ta-rich side, the upper doping limit is $x = 0.4$, while in Nb-rich side, $x = 0.75$, and compositions in between remain as two-phase mixtures. We observe that no matter Ta- or Nb-doping, it does not introduce any additional features to the magnetism compared to the parent compound, however it does modify the existing magnetic transitions in different ways. For instance, doping Nb into $CoTa_2O_6$ gradually lowers $T_N$ of the long-range PM-to-AFM transition, while doping Ta into $CoNb_2O_6$ makes the magnetic transitions more fragile towards external fields with $T_N$ more or less unshifted. The latter one is of a great importance because it may pave a way to strengthen the lower-dimensional quantum phenomena in $CoNb_2O_6$, before the ideal quasi-one-dimensional crystal structure starts to melt down.

## CONFLICT OF INTEREST

The authors declare no conflict of interest.

## ACKNOWLEDGMENTS

This work was supported by the National Natural Science Foundation of China (Grant No. 12504005), the Basic Research Program of Jiangsu (Grant No. BK20251286), the Start-up Research Fund of Southeast University (Grant No. RF1028624196). L.J., S.H. and Y.L. acknowledge the support of Undergraduate Training Programs for Innovation of Jiangsu Province. The authors thank the Center for Fundamental and Interdisciplinary Sciences of Southeast University for the support in structural, magnetization and heat capacity measurements.

**REFERENCES**

[1] L. Balents, Spin liquids in frustrated magnets, Nature **464**, 199 (2010).

[2] Y. Zhou, K. Kanoda, and T.-K. Ng, Quantum spin liquid states, Rev. Mod. Phys. **89**, 025003 (2017).

[3] S. Sachdev, Quantum magnetism and criticality, Nat. Phys. **4**, 173 (2008).

[4] C. Broholm, R. J. Cava, S. A. Kivelson, D. G. Nocera, M. R. Norman, and T. Senthil, Quantum spin liquids, Science **367**, eaay0668 (2020).

[5] Y. Matsuda, T. Shibauchi, and H.-Y. Kee, Kitaev quantum spin liquids, Rev. Mod. Phys. **97**, 045003 (2025).

[6] Q. Faure, S. Takayoshi, S. Petit, V. Simonet, S. Raymond, L.-P. Regnault, M. Boehm, J. S. White, M. Månsson, C. Rüegg, P. Lejay, B. Canals, T. Lorenz, S. C. Furuya, T. Giamarchi, and B. Grenier, Topological quantum phase transition in the ising-like antiferromagnetic spin chain $BaCo_2V_2O_8$, Nat. Phys. **14**, 716 (2018).

[7] S. Agrestini, L. C. Chapon, A. Daoud-Aladine, J. Schefer, A. Gukasov, C. Mazzoli, M. R. Lees, and O. A. Petrenko, Nature of the magnetic order in $Ca_3Co_2O_6$, Phys. Rev. Lett. **101**, 097207 (2008).

[8] J. S. Li, Q. L. Chen, N. Ding, Z. Y. Wu, C. Dong, D. S. Cao, X. Y. Chen, S. H. Zheng, M. F. Liu, L. Huang, B. Yu, W. J. Zhai, P. H. Shi, Y. J. Ma, L. Lin, Z. B. Yan, J. F. Wang, S. Dong, and J.-M. Liu, Successive metamagnetic transitions and magnetoelectric switching in the spin-chain antiferromagnet $CoSe_2O_5$, Phys. Rev. B **111**, 224117 (2025).

[9] G. Lin, J. Jeong, C. Kim, Y. Wang, Q. Huang, T. Masuda, S. Asai, S. Itoh, G. Günther, M. Russina, Z. Lu, J. Sheng, L. Wang, J. Wang, G. Wang, Q. Ren, C. Xi, W. Tong, L. Ling, Z. Liu, L. Wu, J. Mei, Z. Qu, H. Zhou, X. Wang, J.-G. Park, Y. Wan, and J. Ma, Field-induced quantum spin disordered state in spin-1/2 honeycomb magnet $Na_2Co_2TeO_6$, Nat. Commun. **12**, 5559 (2021).

[10] T. Halloran, F. Desrochers, E. Z. Zhang, T. Chen, L. E. Chern, Z. Xu, B. Winn, M. Graves-Brook, M. B. Stone, A. I. Kolesnikov, Y. Qiu, R. Zhong, R. Cava, Y. B. Kim, and C. Broholm, Geometrical frustration versus kitaev interactions in $BaCo_2(AsO_4)_2$, Proc. Natl. Acad. Sci. **120**, e2215509119 (2023).

[11] H. Liu, J. Chaloupka, and G. Khaliullin, Kitaev spin liquid in 3*d* transition metal compounds, Phys. Rev. Lett. **125**, 047201 (2020).

[12] R. Zhong, T. Gao, N. P. Ong, and R. J. Cava, Weak-field induced nonmagnetic state in a Co-based honeycomb, Sci. Adv. **6**, eaay6953 (2020).

[13] T. Liang, S. M. Koohpayeh, J. W. Krizan, T. M. McQueen, R. J. Cava, and N. P. Ong, Heat capacity peak at the quantum critical point of the transverse Ising magnet $CoNb_2O_6$, Nat. Commun. **6**, 7611 (2015).

[14] S. Lee, R. K. Kaul, and L. Balents, Interplay of quantum criticality and geometric frustration in columbite, Nat. Phys. **6**, 702 (2010).
[15] L. Jin, S. Peng, A. N. Rutherford, X. Xu, D. Ni, C. Yang, Y. J. Byeon, W. Xie, H. Zhou, X. Dai, and R. J. Cava, A pyroxene-based quantum magnet with multiple magnetization plateaus, Sci. Adv. **10**, eadp4685 (2024).
[16] S. Kimura, H. Yashiro, K. Okunishi, M. Hagiwara, Z. He, K. Kindo, T. Taniyama, and M. Itoh, Field-induced order-disorder transition in antiferromagnetic $BaCo_2V_2O_8$ driven by a softening of spinon excitation, Phys. Rev. Lett. **99**, 087602 (2007).
[17] A. K. Bera, B. Lake, F. H. L. Essler, L. Vanderstraeten, C. Hubig, U. Schollwöck, A. T. M. N. Islam, A. Schneidewind, and D. L. Quintero-Castro, Spinon confinement in a quasi-one-dimensional anisotropic heisenberg magnet, Phys. Rev. B **96**, 054423 (2017).
[18] N. J. Robinson, F. H. L. Essler, I. Cabrera, and R. Coldea, Quasiparticle breakdown in the quasi-one-dimensional ising ferromagnet $CoNb_2O_6$, Phys. Rev. B **90**, 174406 (2014).
[19] C. M. Morris, R. Valdés Aguilar, A. Ghosh, S. M. Koohpayeh, J. Krizan, R. J. Cava, O. Tchernyshyov, T. M. McQueen, and N. P. Armitage, Hierarchy of bound states in the one-dimensional ferromagnetic ising chain $CoNb_2O_6$ investigated by high-resolution time-domain terahertz spectroscopy, Phys. Rev. Lett. **112**, 137403 (2014).
[20] L. Jin, and R. J. Cava, Recent Progress in Studies of Cobalt-based Quasi-1-dimensional Quantum Magnets, Front. Phys. **20**, 34301 (2025).
[21] A. W. Kinross, M. Fu, T. J. Munsie, H. A. Dabkowska, G. M. Luke, S. Sachdev, and T. Imai, Evolution of quantum fluctuations near the quantum critical point of the transverse field ising chain system $CoNb_2O_6$, Phys. Rev. X **4**, 031008 (2014).
[22] R. Coldea, D. A. Tennant, E. M. Wheeler, E. Wawrzynska, D. Prabhakaran, M. Telling, K. Habicht, P. Smeibidl, and K. Kiefer, Quantum criticality in an Ising chain: experimental evidence for emergent $E_8$ symmetry, Science **327**, 177 (2010).
[23] J. N. Reimers, J. E. Greedan, C. V. Stager, and R. Kremer, Crystal structure and magnetism in $CoSb_2O_6$ and $CoTa_2O_6$, J. Solid State Chem. **83**, 20 (1989).
[24] A. B. Christian, A. T. Schye, K. O. White, and J. J. Neumeier, Magnetic, thermal, and optical properties of single-crystalline $CoTa_2O_6$ and $FeTa_2O_6$ and their anisotropic magnetocaloric effect, J. Phys.: Condens. Matter **30**, 195803 (2018).
[25] I. S. Mulla, N. Natarajan, A. B. Gaikwad, V. Samuel, U. N. Guptha, and V. Ravi, A coprecipitation technique to prepare $CoTa_2O_6$ and $CoNb_2O_6$, Mater. Lett. **61**, 2127 (2007).
[26] E. J. Kinast, C. A. dos Santos, D. Schmitt, O. Isnard, M. A. Gusmão, and J. B. M. da Cunha, Magnetic structure of the quasi-two-dimensional compound $CoTa_2O_6$, J. Alloys Compd. **491**, 41 (2010).
[27] C. Heid, H. Weitzel, P. Burlet, M. Bonnet, W. Gonschorek, T. Vogt, J. Norwig, and H. Fuess, Magnetic phase diagram of $CoNb_2O_6$: a neutron diffraction study, J. Magn. Magn. Mater. **151**, 123 (1995).

[28] S. Kobayashi, S. Mitsuda, M. Ishikawa, K. Miyatani, and K. Kohn, Three-dimensional magnetic ordering in the quasi-one-dimensional ising magnet $CoNb_2O_6$ with partially released geometrical frustration, Phys. Rev. B **60**, 3331 (1999).
[29] W. Jia, Z. Huang, and L. Li, Ta doping effect on microstructure and microwave dielectric properties of $CoNb_2O_6$-based ceramics and mechanism study, J. Am. Ceram. Soc. **108**, e20226 (2025).
[30] R. D. Shannon, Revised effective ionic radii and systematic studies of interatomic distances in halides and chalcogenides, Acta Crystallogr. A **32**, 751 (1976).
[31] S. Thota, S. Ghosh, Maruthi R, D. C. Joshi, R. Medwal, R. S. Rawat, and M. S. Seehra, Magnetic ground state and exchange interactions in the ising chain ferromagnet $CoNb_2O_6$, Phys. Rev. B **103**, 064415 (2021).
[32] W. Scharf, H. Weitzel, I. Yaeger, I. Maartense, and B. M. Wanklyn, Magnetic structures of $CoNb_2O_6$, J. Magn. Magn. Mater. **13**, 121 (1979).
[33] S. Kobayashi, S. Mitsuda, K. Hosoya, H. Yoshizawa, T. Hanawa, M. Ishikawa, K. Miyatani, K. Saito, and K. Kohn, Competition between the inter-chain interaction and single-ion anisotropy in $CoNb_2O_6$, Physica B **213–214**, 176 (1995).
[34] T. Hanawa, K. Shinkawa, M. Ishikawa, K. Miyatani, K. Saito, and K. Kohn, Anisotropic specific heat of $CoNb_2O_6$ in magnetic fields, J. Phys. Soc. Jpn. **63**, 2706 (1994).
[35] J. H. Lee, M. Kratochvílová, H. Cao, Z. Yamani, J. S. Kim, J.-G. Park, G. R. Stewart, and Y. S. Oh, Unconventional critical behavior in the quasi-one-dimensional $S = 1$ chain $NiTe_2O_5$, Phys. Rev. B **100**, 144441 (2019).
[36] L. J. de Jongh, Experiments on simple magnetic model systems, J. Appl. Phys. **49**, 1305 (1978).

## Supplementary Information

# The Structure-tuning Magnetism in the $Co(Nb_xTa_{1-x})_2O_6$ Series

Sirui Huang[1, #], Lun Jin[1, #, *], Yuchen Liu[1], Xiyu Chen[1] and Leili Tan[1]

[1]Key Laboratory of Quantum Materials and Devices of Ministry of Education, School of Physics, Southeast University, Nanjing 211189, China

# These authors contributed equally to this work.

* E-mails of corresponding authors: ljin@seu.edu.cn

## Table of Content

## I. Structural Characterization

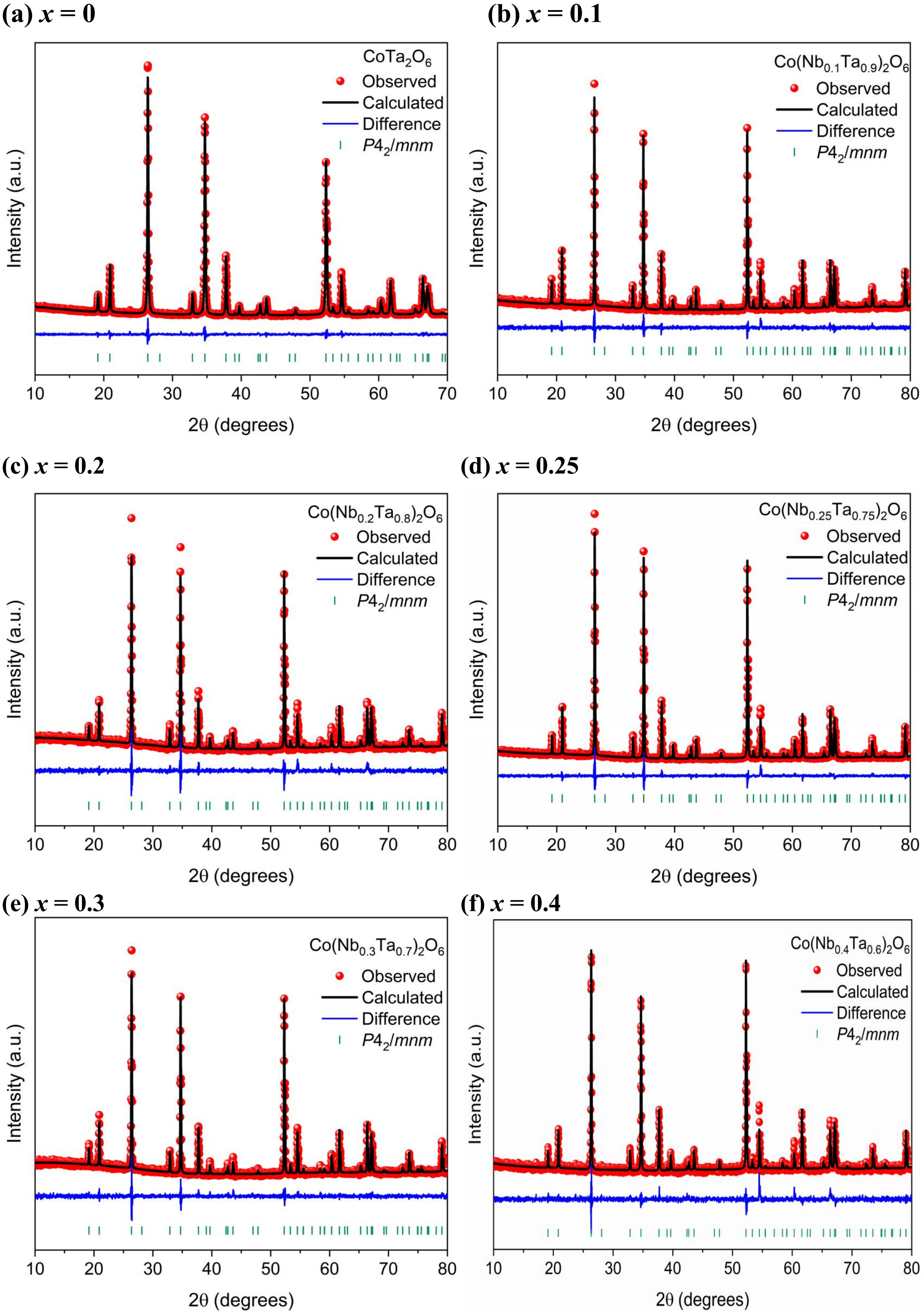

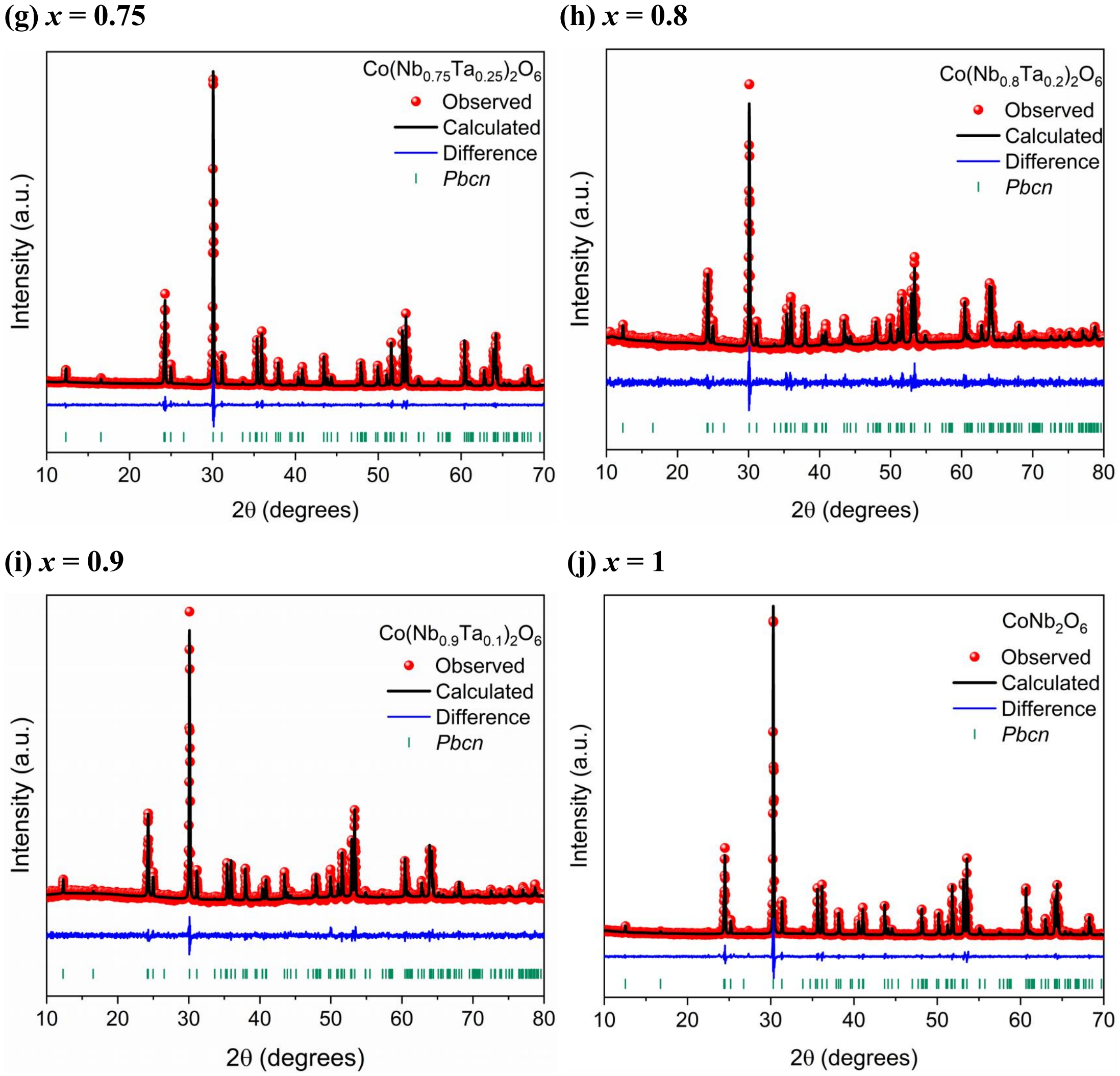

FIG. S1. Rietveld refinement profiles of the powder X-ray diffraction patterns for $Co(Nb_xTa_{1-x})_2O_6$ ($0 \leq x \leq 1$). Red circles, black lines, blue lines, and green ticks denote the observed intensities, calculated intensities, differences, and Bragg reflections, respectively.

TABLE S1. Lattice parameters refined from the powder X-ray diffraction patterns for $Co(Nb_xTa_{1-x})_2O_6$ ($0 \leq x \leq 1$).

| *x* | Space group | *a* (Å) | *b* (Å) | *c* (Å) |
|---|---|---|---|---|
| 0 | $P4_2/mnm$ | 4.736(2) | 4.736(2) | 9.173(1) |
| 0.1 | $P4_2/mnm$ | 4.737(4) | 4.737(4) | 9.171(2) |
| 0.2 | $P4_2/mnm$ | 4.737(4) | 4.737(4) | 9.168(2) |
| 0.25 | $P4_2/mnm$ | 4.738(2) | 4.738(2) | 9.169(1) |
| 0.3 | $P4_2/mnm$ | 4.739(1) | 4.739(1) | 9.168(2) |
| 0.4 | $P4_2/mnm$ | 4.739(3) | 4.739(3) | 9.164(3) |
| two-phase mixture | | | | |
| 0.75 | *Pbcn* | 14.154(1) | 5.710(2) | 5.047(2) |
| 0.8 | *Pbcn* | 14.145(2) | 5.707(3) | 5.045(1) |
| 0.9 | *Pbcn* | 14.151(2) | 5.704(3) | 5.044(1) |
| 1 | *Pbcn* | 14.152(2) | 5.710(1) | 5.044(5) |

# II. Magnetization and Curie-Weiss Fits

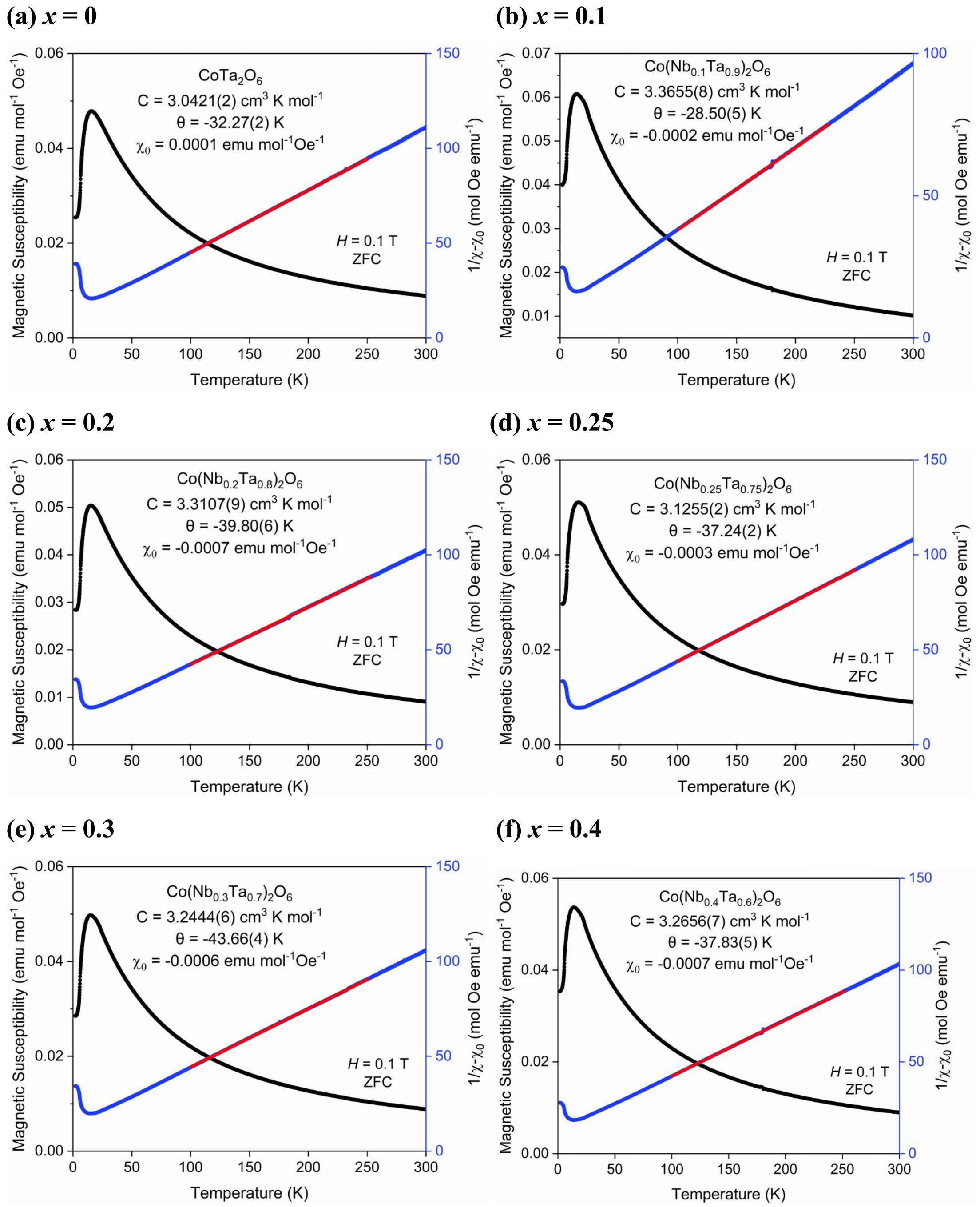

FIG. S2. Temperature-dependent magnetic susceptibility and Curie-Weiss fitting plots for the Ta-rich compositions. The red fitting segments indicate the temperature windows used for Curie-Weiss analysis.

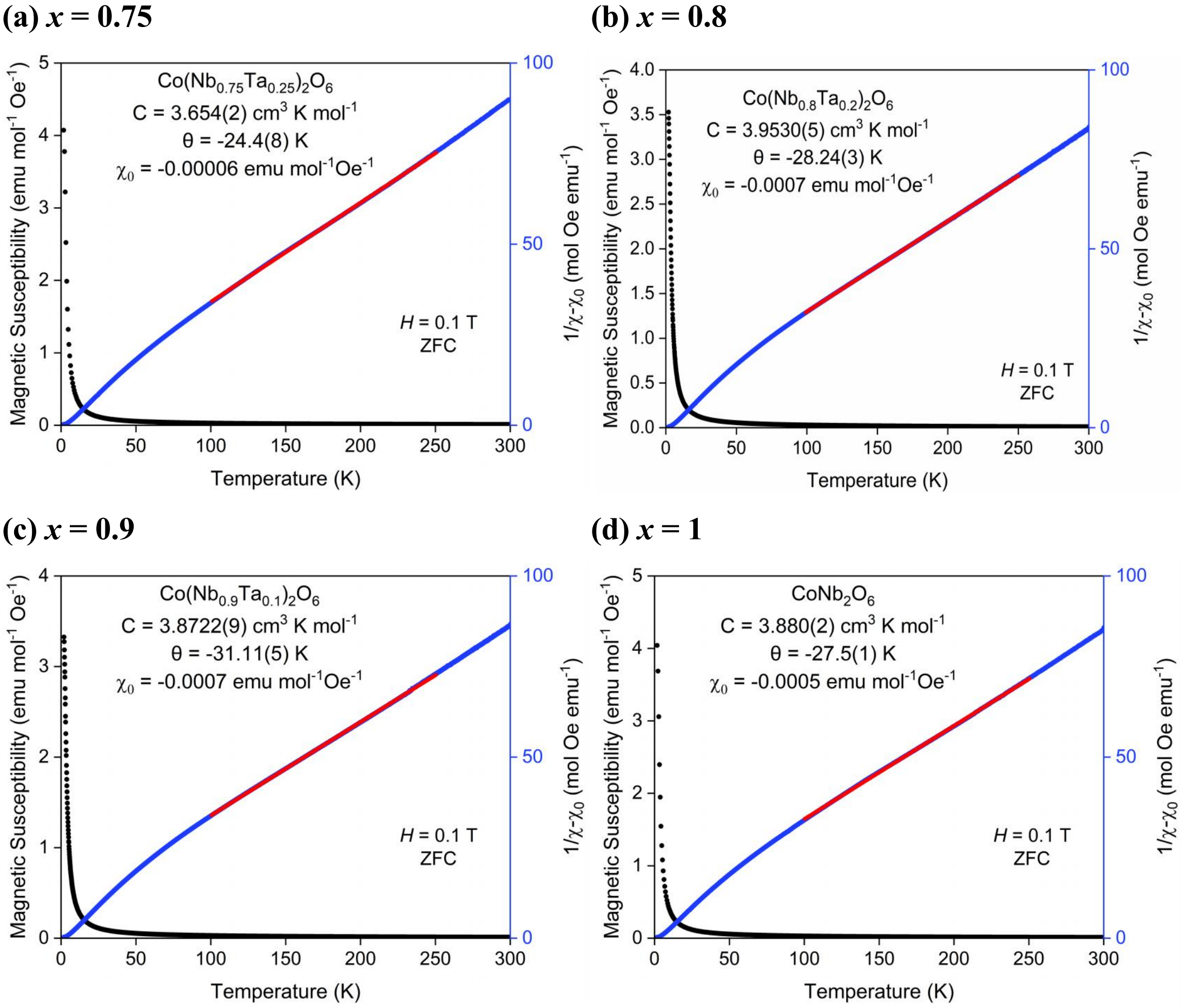

FIG. S3. Temperature-dependent magnetic susceptibility and Curie-Weiss fitting plots for the Nb-rich compositions. The red fitting segments indicate the temperature windows used for Curie-Weiss analysis.

TABLE S2. Curie-Weiss fitting parameters.

| $x$ | $C$ (cm$^3$ K mol$^{-1}$) | $\theta$ (K) | $\mu_{eff}$ ($\mu_B$/f.u.) |
|---|---|---|---|
| 0 | 3.0421(2) | -32.27(2) | 4.93 |
| 0.1 | 3.3655(8) | -28.50(5) | 5.19 |
| 0.2 | 3.3107(9) | -39.80(6) | 5.15 |
| 0.25 | 3.1255(2) | -37.24(2) | 5.00 |
| 0.3 | 3.2444(6) | -43.66(4) | 5.09 |
| 0.4 | 3.2656(7) | -37.83(5) | 5.11 |
| 0.75 | 3.654(2) | -24.4(8) | 5.41 |
| 0.8 | 3.9530(5) | -28.24(3) | 5.62 |
| 0.9 | 3.8722(9) | -31.11(5) | 5.57 |
| 1 | 3.880(2) | -27.5(1) | 5.57 |

## III. Field-sensitivity of the Magnetic Susceptibility

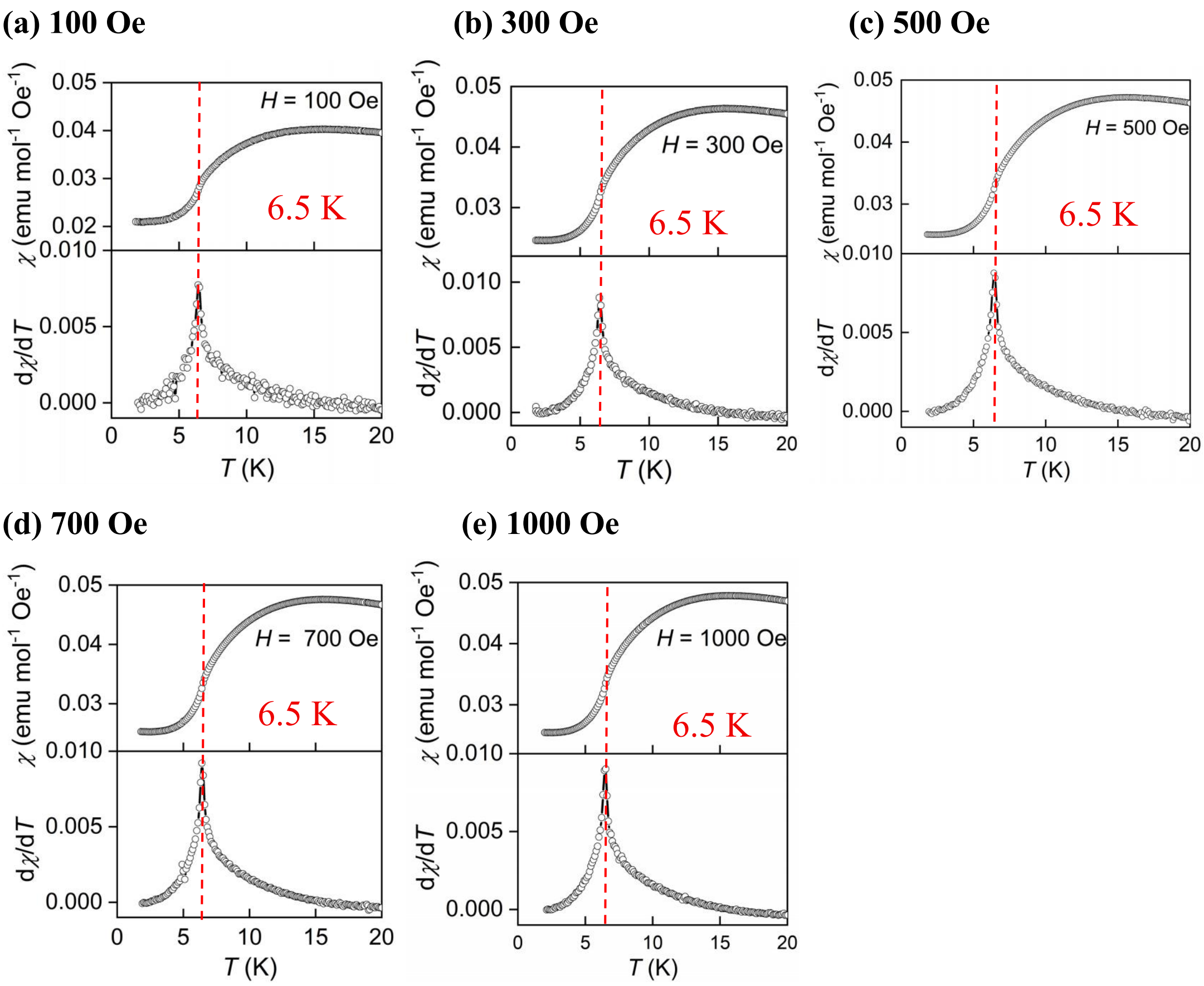

FIG. S4. Low-temperature zero-field-cooled magnetic susceptibility and derivative analysis for $CoTa_2O_6$ ($x = 0$) under applied fields of 100, 300, 500, 700, and 1000 Oe.

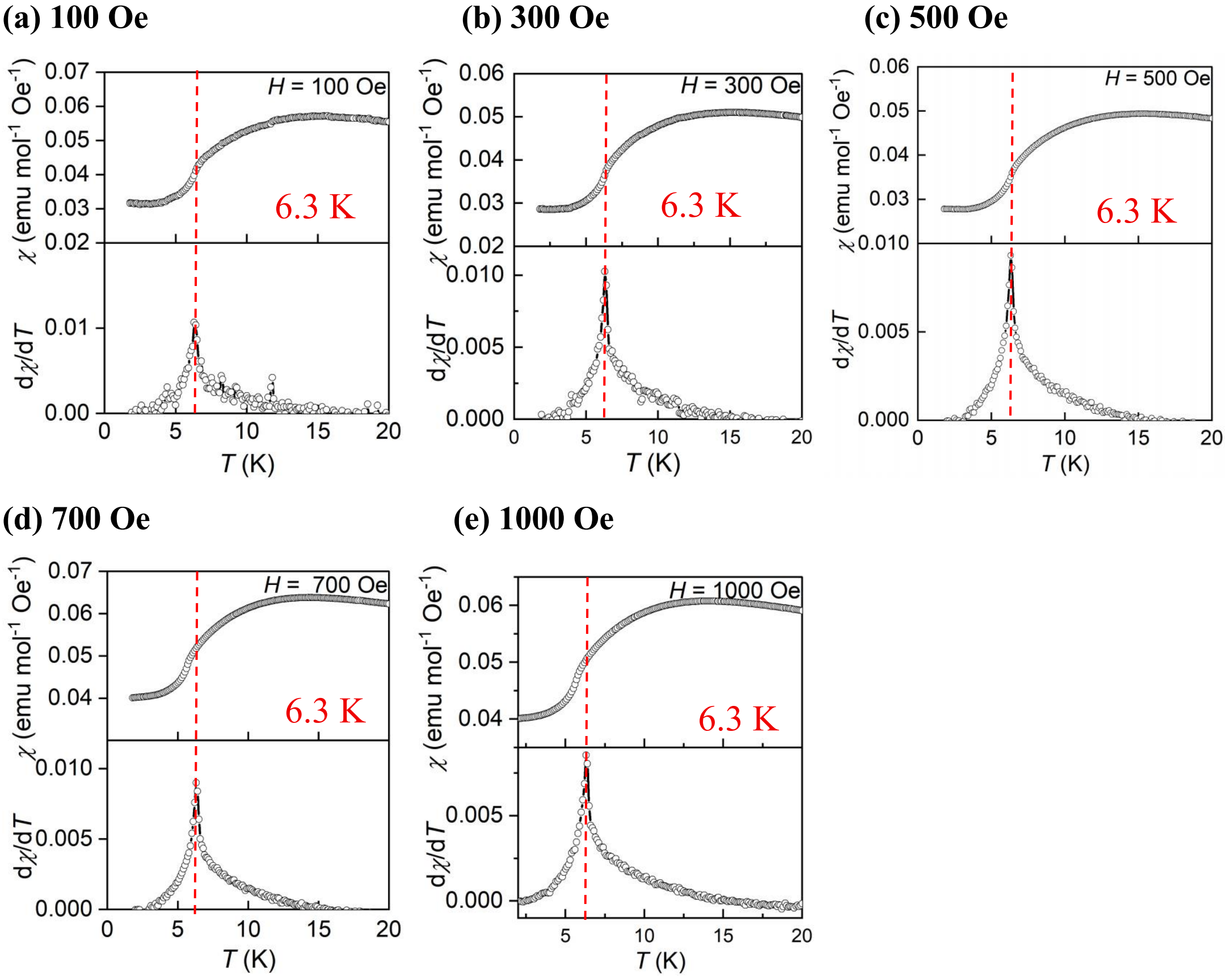

FIG. S5. Low-temperature zero-field-cooled magnetic susceptibility and derivative analysis for $Co(Nb_{0.1}Ta_{0.9})_2O_6$ ($x = 0.1$) under applied fields of 100, 300, 500, 700, and 1000 Oe.

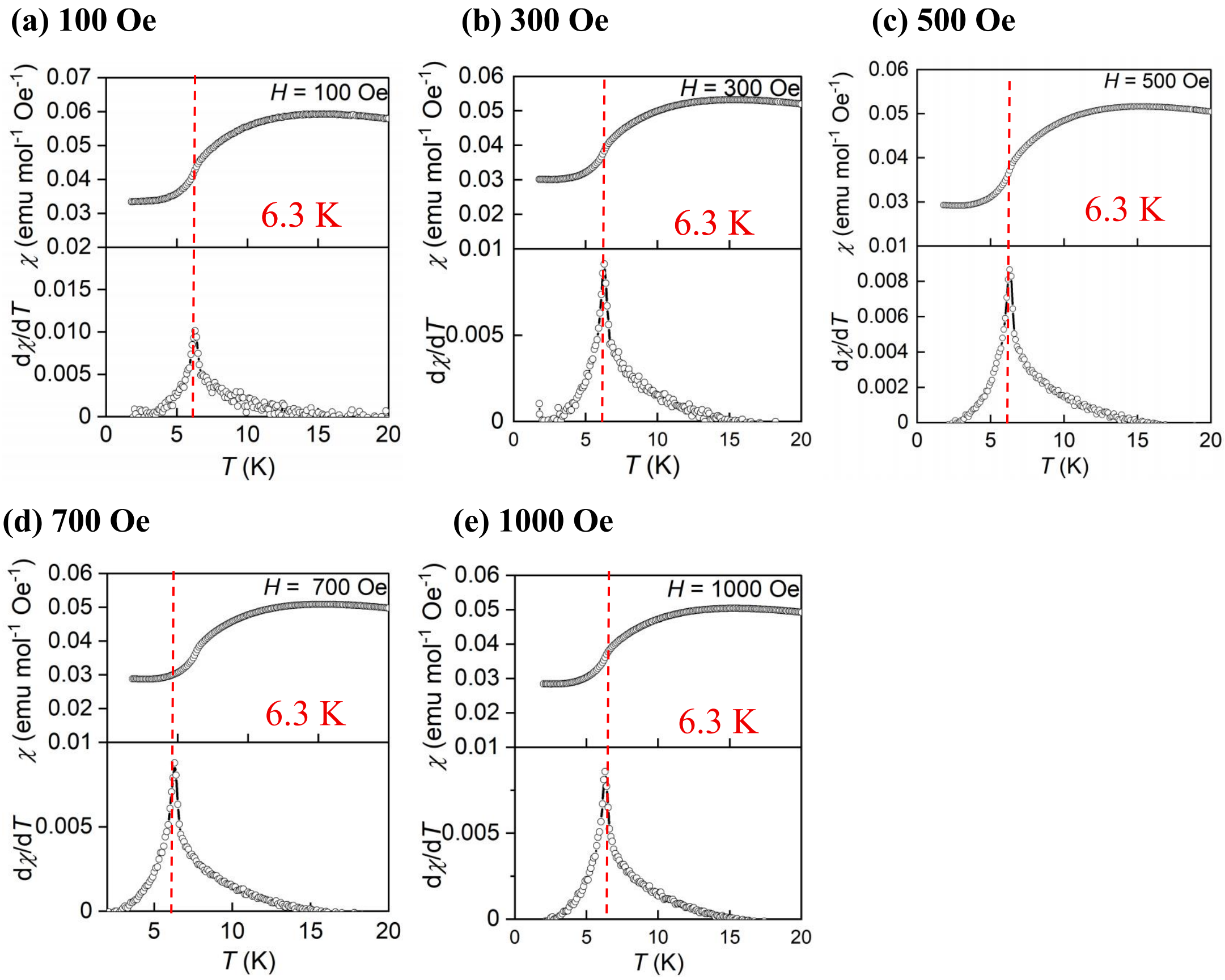

FIG. S6. Low-temperature zero-field-cooled magnetic susceptibility and derivative analysis for $Co(Nb_{0.2}Ta_{0.8})_2O_6$ ($x$ = 0.2) under applied fields of 100, 300, 500, 700, and 1000 Oe.

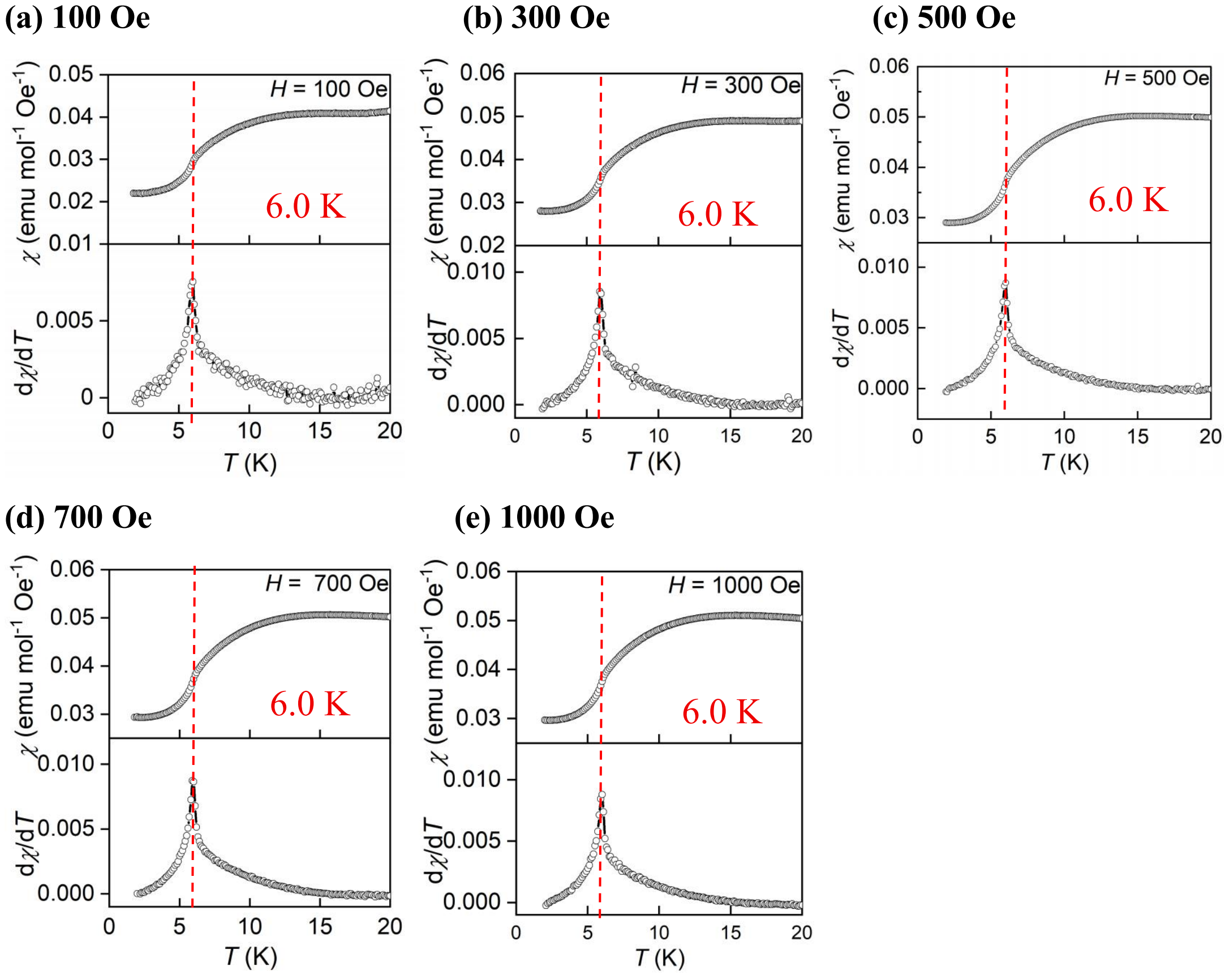

FIG. S7. Low-temperature zero-field-cooled magnetic susceptibility and derivative analysis for $Co(Nb_{0.25}Ta_{0.75})_2O_6$ ($x$ = 0.25) under applied fields of 100, 300, 500, 700, and 1000 Oe.

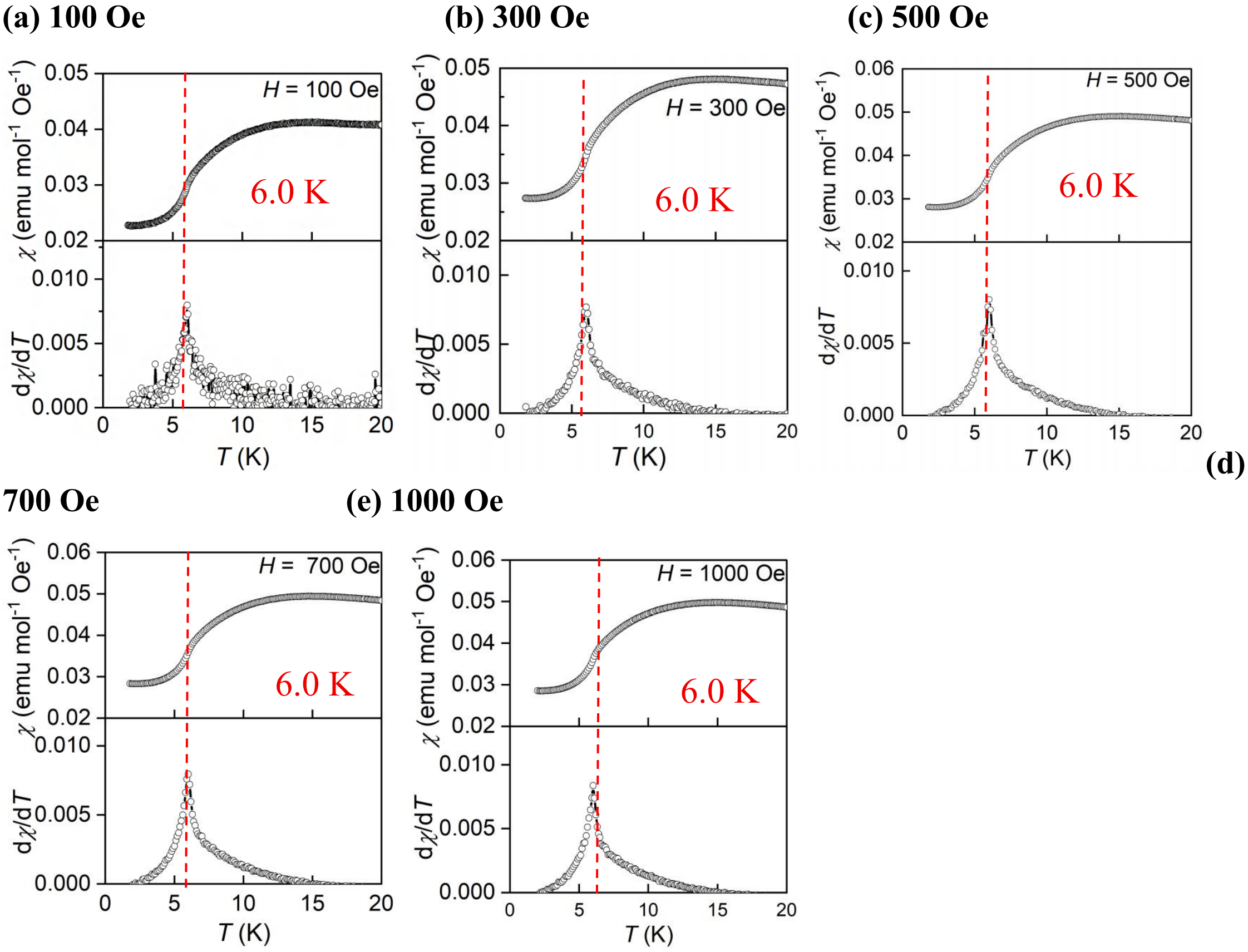

FIG. S8. Low-temperature zero-field-cooled magnetic susceptibility and derivative analysis for $Co(Nb_{0.3}Ta_{0.7})_2O_6$ ($x = 0.3$) under applied fields of 100, 300, 500, 700, and 1000 Oe.

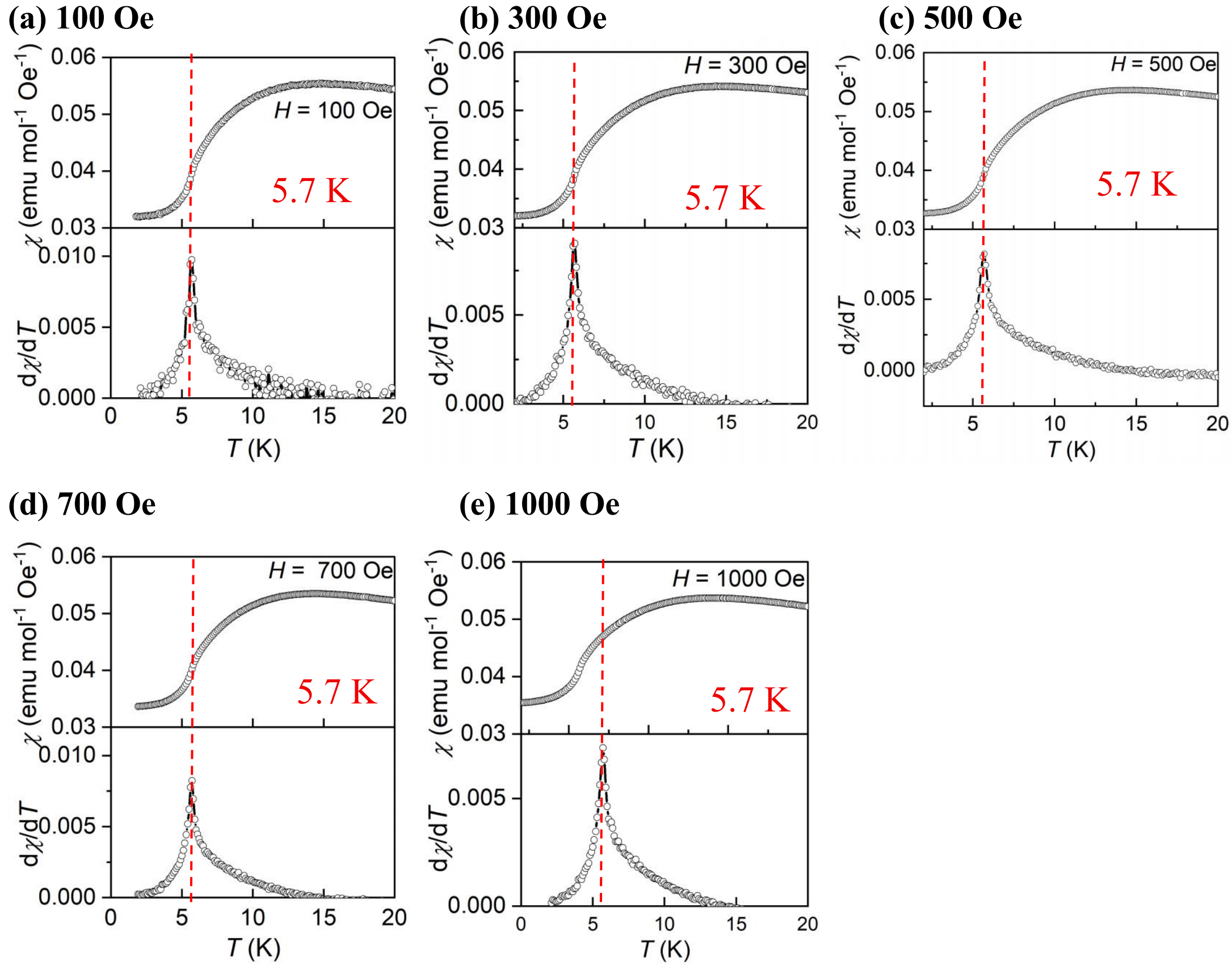

FIG. S9. Low-temperature zero-field-cooled magnetic susceptibility and derivative analysis for $Co(Nb_{0.4}Ta_{0.6})_2O_6$ ($x$ = 0.4) under applied fields of 100, 300, 500, 700, and 1000 Oe.

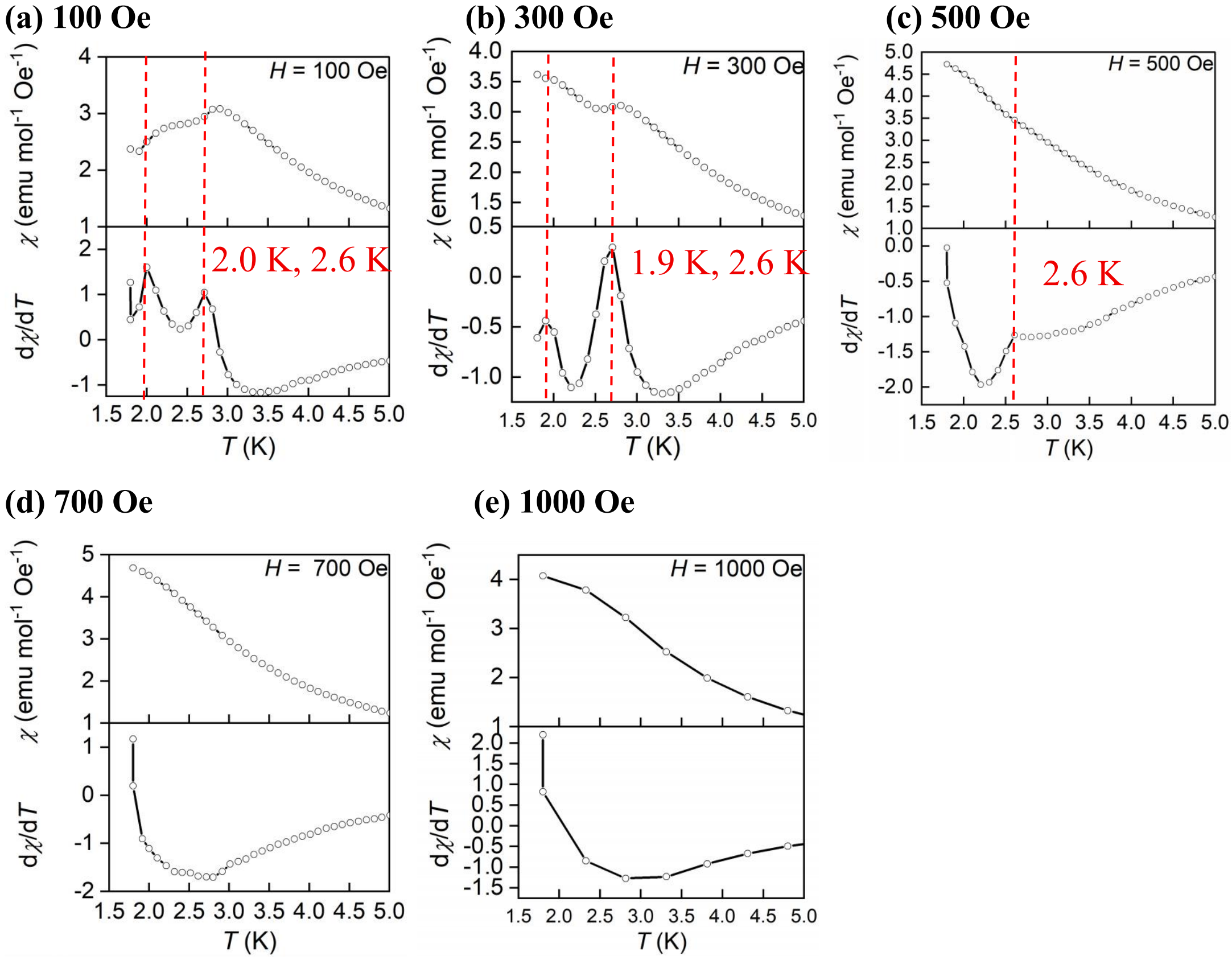

FIG. S10. Low-temperature zero-field-cooled magnetic susceptibility and derivative analysis for $Co(Nb_{0.75}Ta_{0.25})_2O_6$ ($x$ = 0.75) under applied fields of 100, 300, 500, 700, and 1000 Oe.

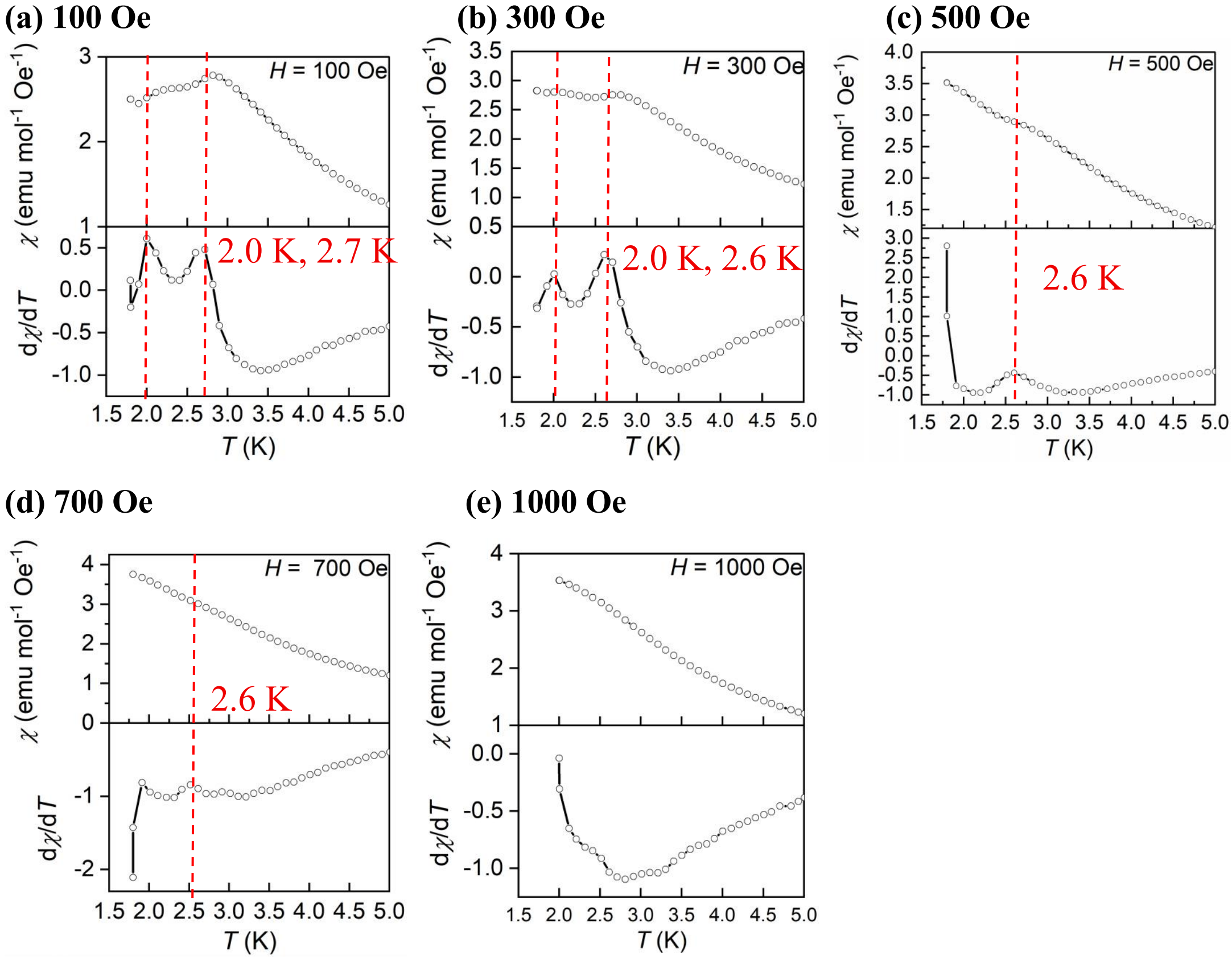

FIG. S11. Low-temperature zero-field-cooled magnetic susceptibility and derivative analysis for $Co(Nb_{0.8}Ta_{0.2})_2O_6$ ($x$ = 0.8) under applied fields of 100, 300, 500, 700, and 1000 Oe.

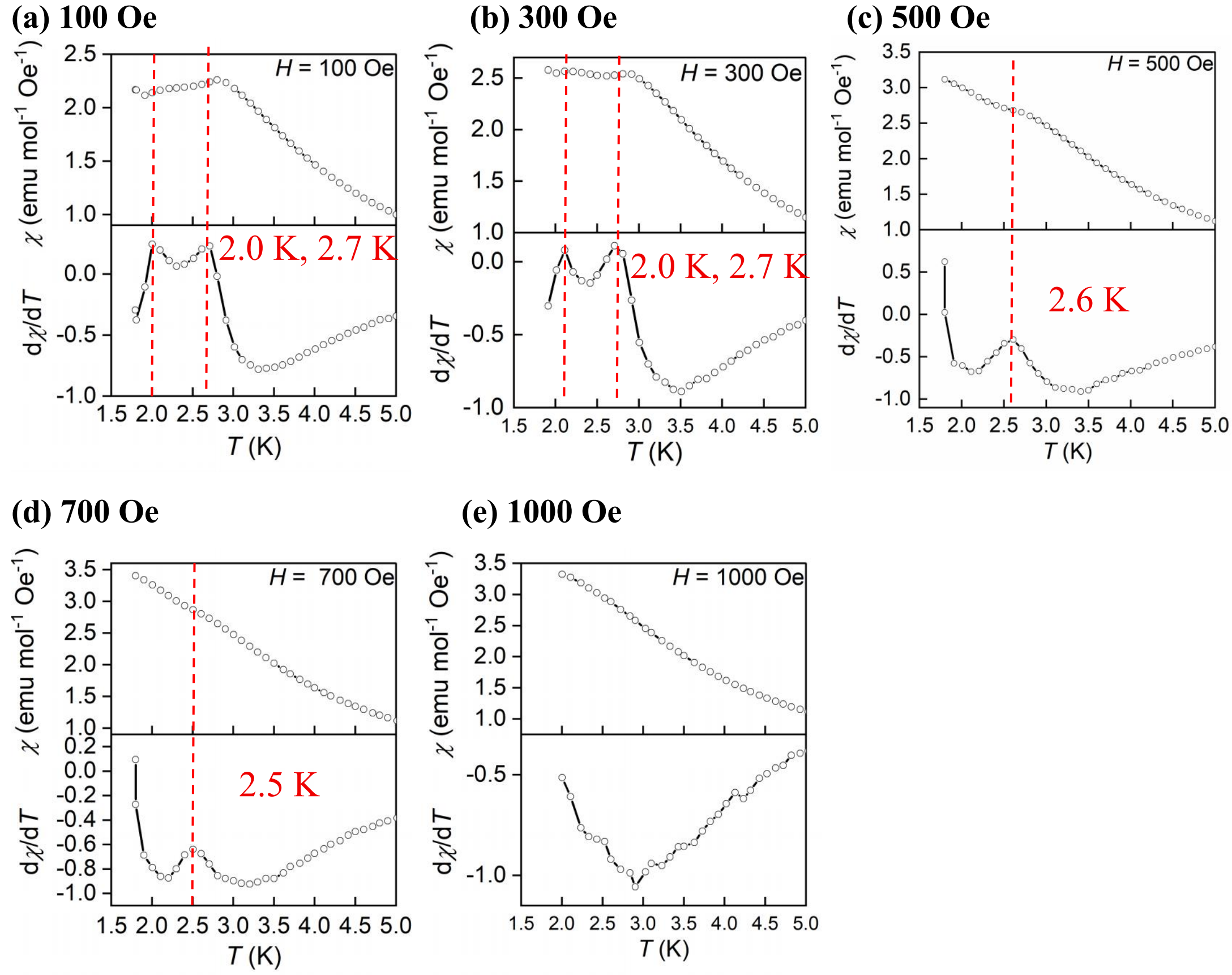

FIG. S12. Low-temperature zero-field-cooled magnetic susceptibility and derivative analysis for $Co(Nb_{0.9}Ta_{0.1})_2O_6$ ($x = 0.9$) under applied fields of 100, 300, 500, 700, and 1000 Oe.

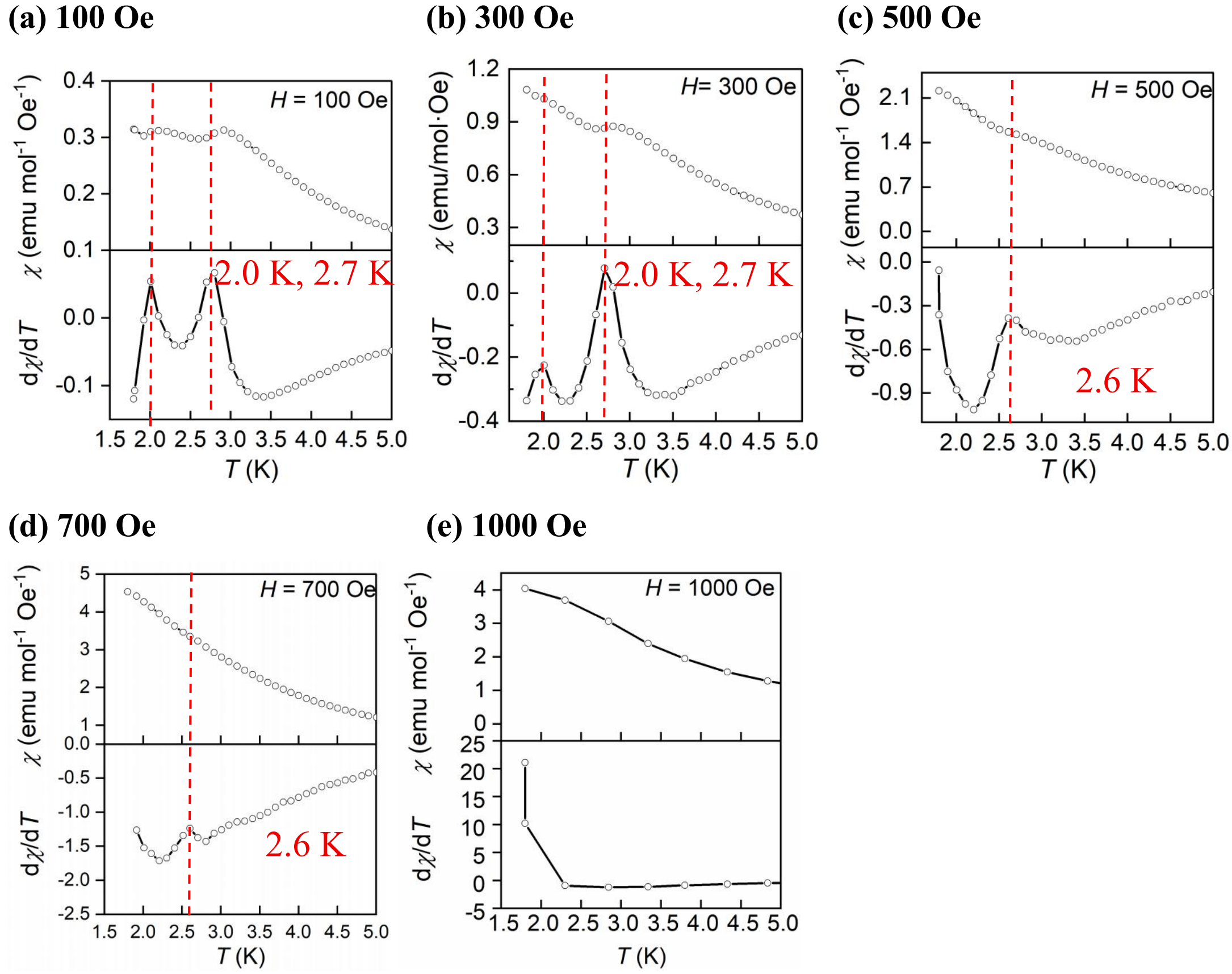

FIG. S13. Low-temperature zero-field-cooled magnetic susceptibility and derivative analysis for $CoNb_2O_6$ ($x = 1$) under applied fields of 100, 300, 500, 700, and 1000 Oe.

## IV. Heat Capacity and Magnetic Entropy

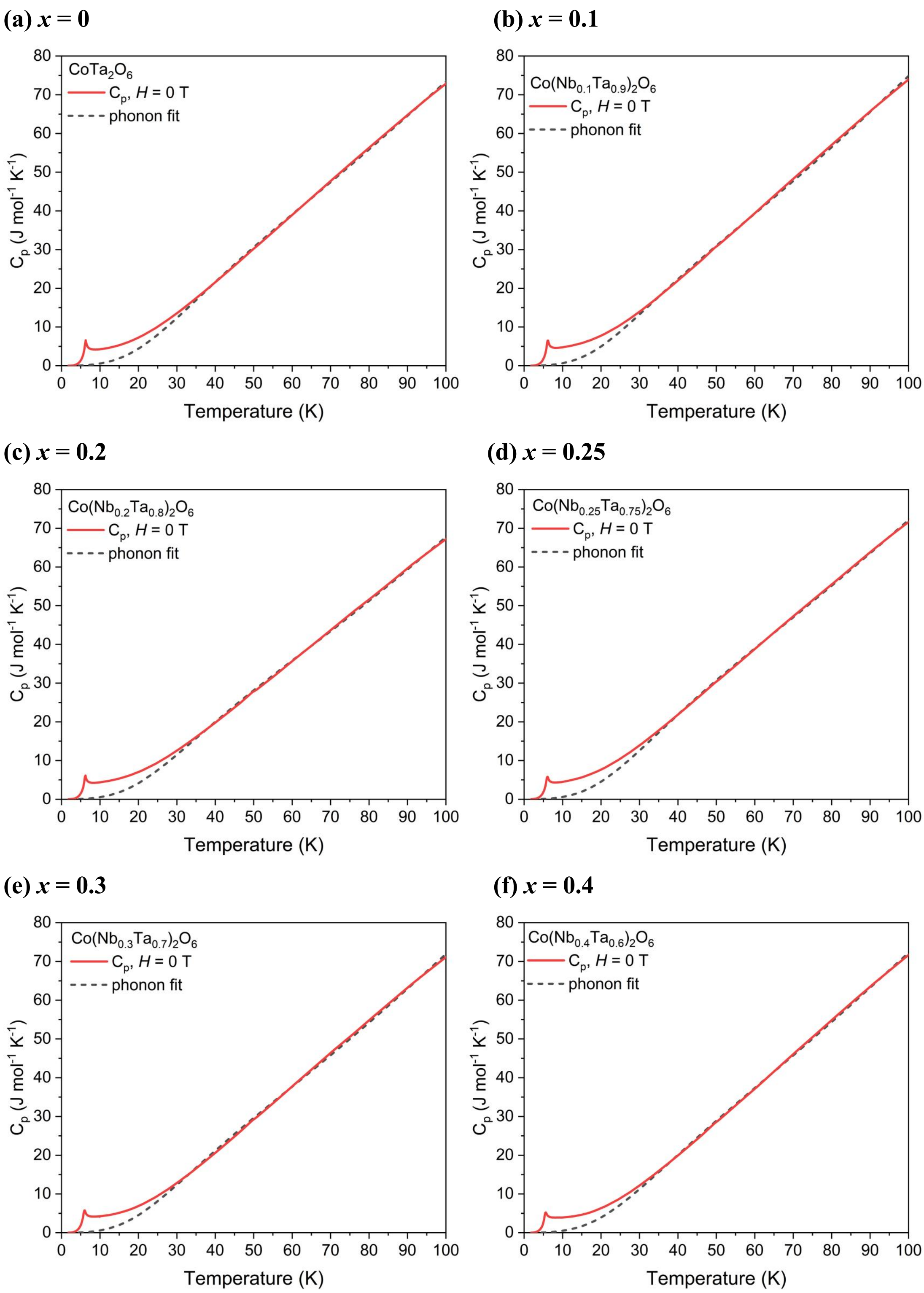

FIG. S14. Zero-field heat capacity Cp and fitted lattice contribution for Ta-rich compositions. The phonon background was fitted in the high-temperature region described in the main text.

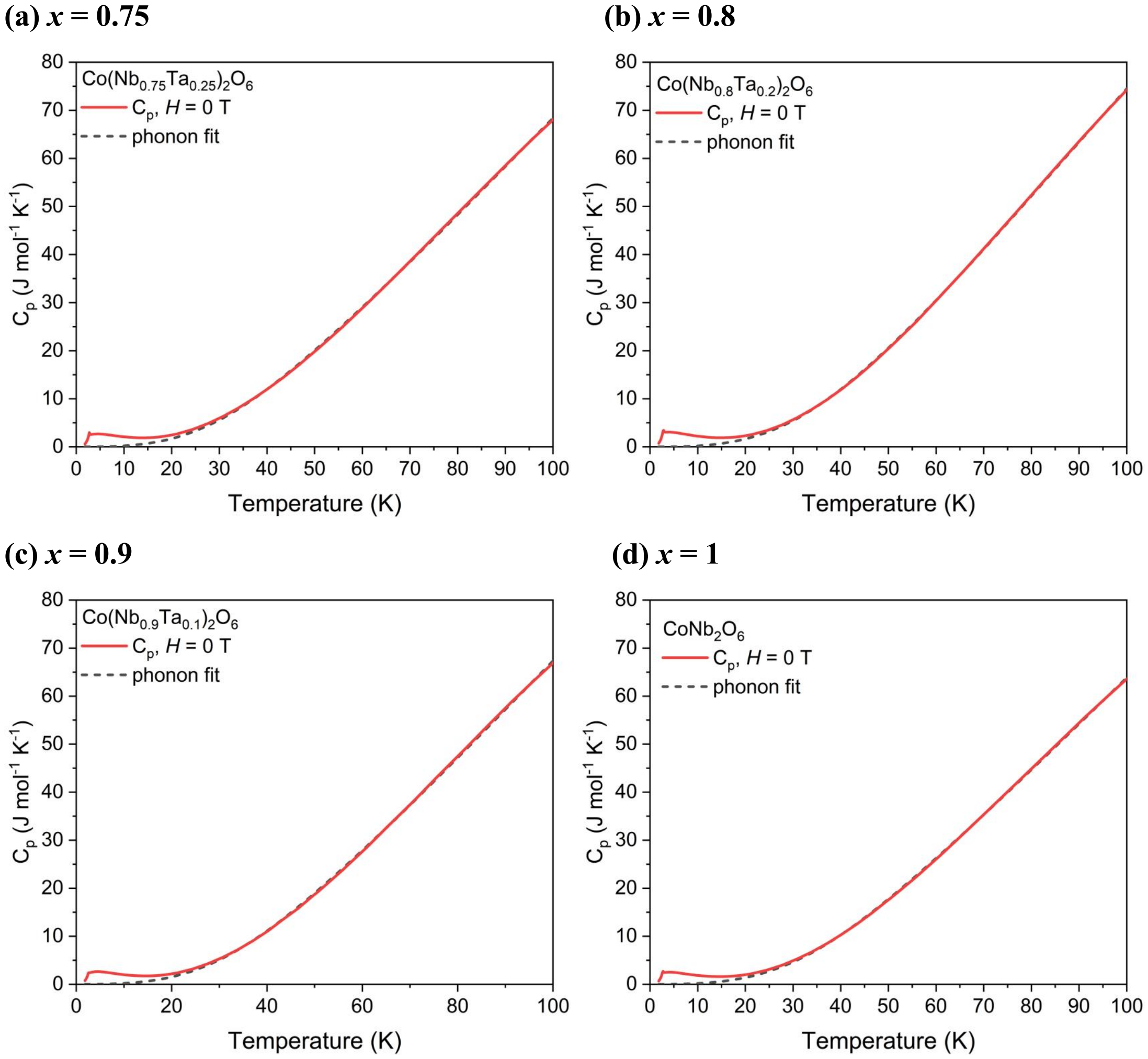

FIG. S15. Zero-field heat capacity Cp and fitted lattice contribution for Nb-rich compositions. The phonon background was fitted in the high-temperature region described in the main text.

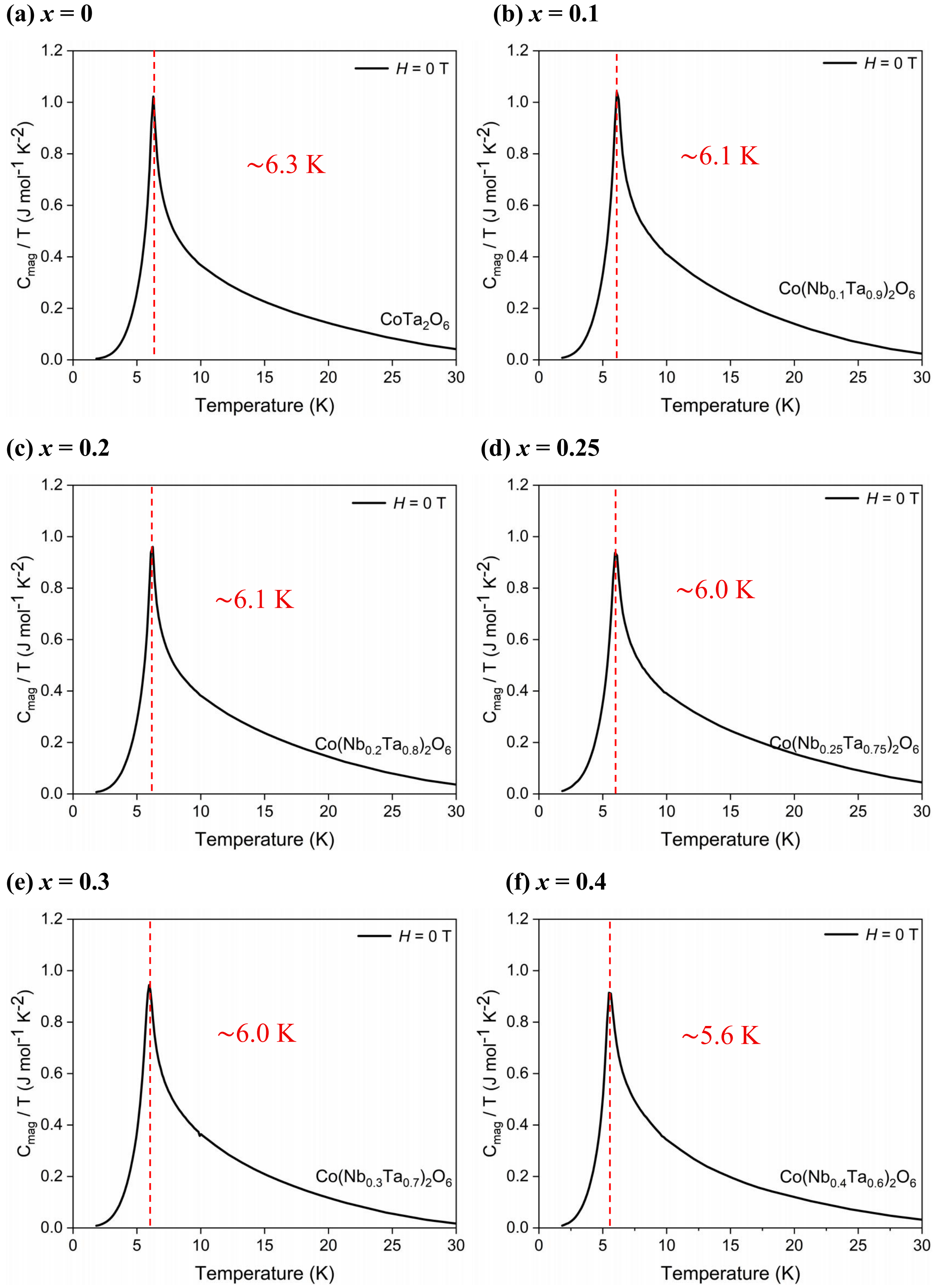

FIG. S16. Magnetic contribution $C_{mag}/T$ after subtraction of the fitted lattice background for Ta-rich compositions.

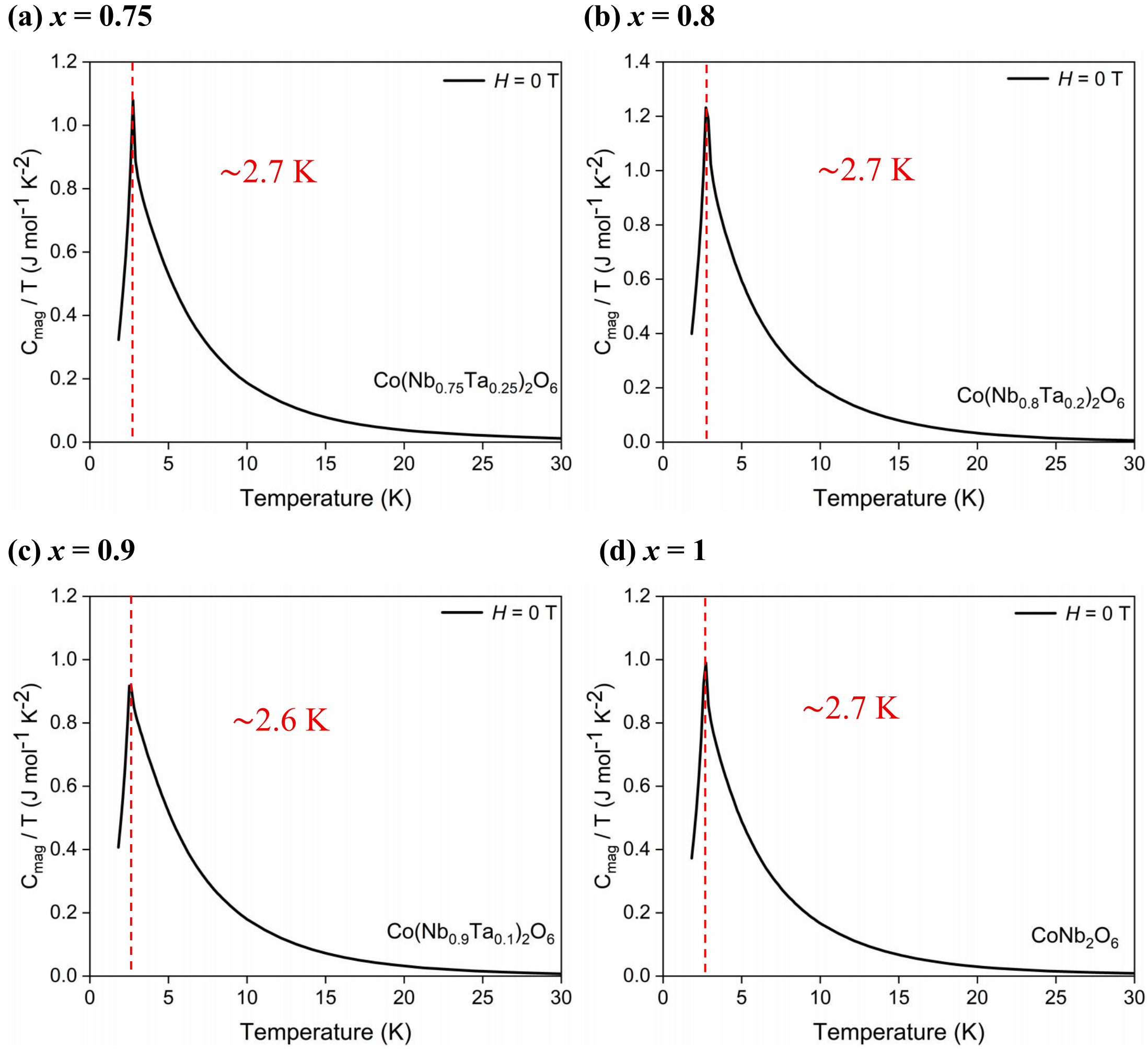

FIG. S17. Magnetic contribution $C_{mag}/T$ after subtraction of the fitted lattice background for Nb-rich compositions.

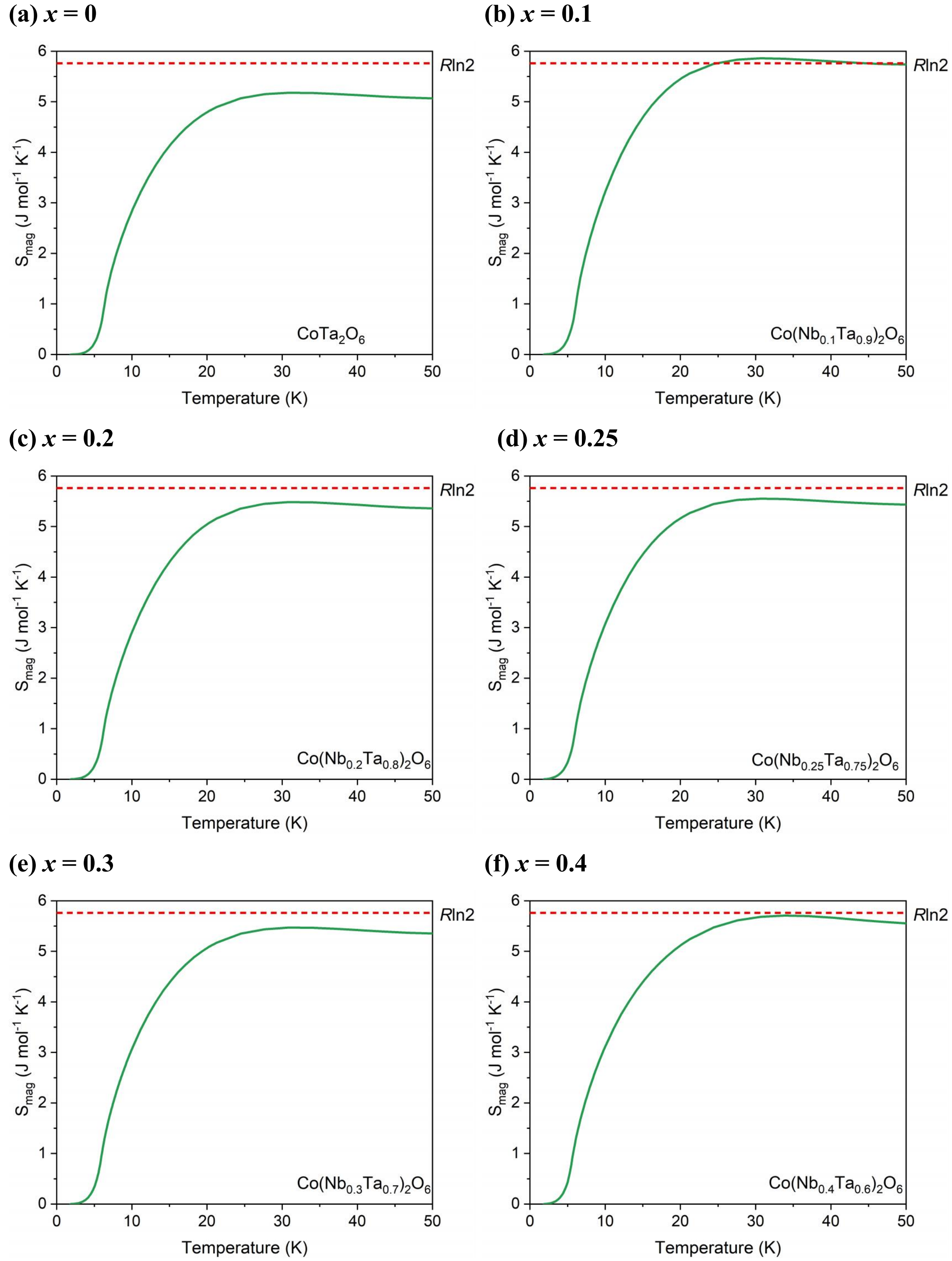

FIG. S18. Magnetic entropy $S_{mag}$ obtained by integrating $C_{mag}/T$ for Ta-rich compositions.

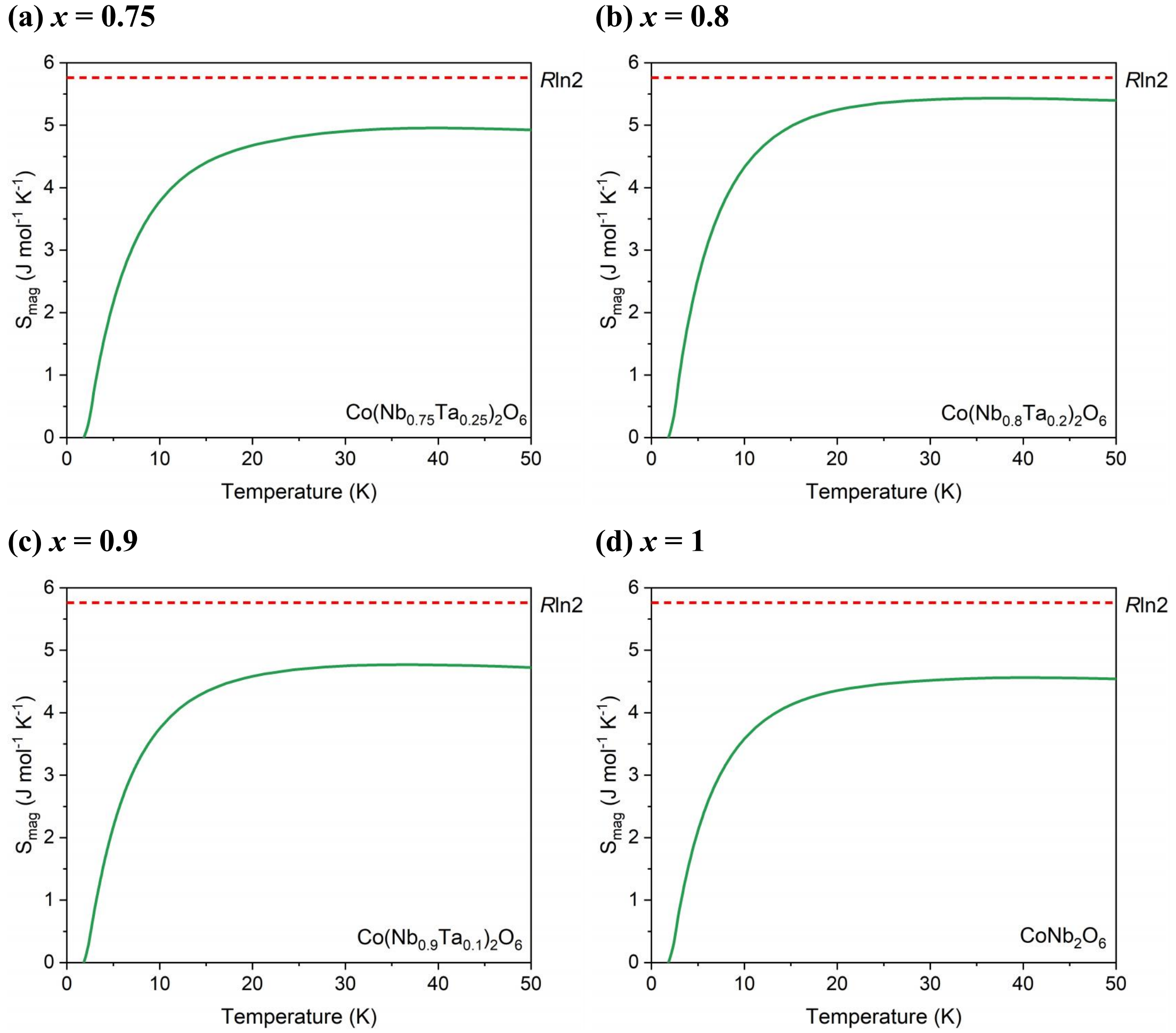

FIG. S19. Magnetic entropy $S_{mag}$ obtained by integrating $C_{mag}/T$ for Nb-rich compositions.

TABLE S3. Heat-capacity fitting and entropy summary.

| $x$ | Fit range (K) | $\theta_{D1}$ (K) | $\theta_{D2}$ (K) | $\theta_{D3}$ (K) | $S_{mag}$(J mol$^{-1}$ K$^{-1}$) |
|---|---|---|---|---|---|
| 0 | 25-100 | 151.8 | 366.1 | 789.2 | 5.12 |
| 0.1 | 27-100 | 148.2 | 367.5 | 774.1 | 5.78 |
| 0.2 | 25-100 | 155.7 | 402.8 | 837.3 | 5.40 |
| 0.25 | 25-100 | 147.6 | 368.2 | 808.6 | 5.48 |
| 0.3 | 25-100 | 157.2 | 373.2 | 805.1 | 5.40 |
| 0.4 | 30-100 | 205.5 | 627.9 | 739.7 | 5.61 |
| 0.75 | 25-100 | 284.0 | 527.2 | 736.9 | 4.91 |
| 0.8 | 20-100 | 296.0 | 474.7 | 675.8 | 5.39 |
| 0.9 | 25-100 | 301.1 | 466.7 | 755.4 | 4.72 |
| 1 | 20-100 | 306.8 | 490.9 | 785.0 | 4.53 |